%% file: ms.tex
\documentclass[
  journal=pasa,
  manuscript=research-paper, %% or "review"
  year=2022,
  volume=39,
]{cup-journal}

\usepackage{microtype,siunitx,booktabs}
\usepackage{aas-macros}
\usepackage{epstopdf}
\usepackage{caption}
\usepackage{subcaption}

\usepackage{amsmath,amssymb}
\usepackage{gensymb}
\usepackage{longtable,threeparttablex}
\usepackage{physics}
\usepackage{comment}
\usepackage{hyperref} % Hyperlinks

\hypersetup{colorlinks,citecolor=blue,linkcolor=blue,urlcolor=blue}
\usepackage{listings}
\definecolor{codegreen}{rgb}{0,0.6,0}
\definecolor{codered}{rgb}{0.6,0,0}
\definecolor{codeblue}{rgb}{0,0,0.6}
\definecolor{codegray}{rgb}{0.5,0.5,0.5}
\definecolor{codepurple}{rgb}{0.58,0,0.82}
\definecolor{codeorange}{RGB}{200, 112, 60}
\definecolor{backcolour}{rgb}{0.95,0.95,0.95}

\lstdefinestyle{mystyle}{
    backgroundcolor=\color{backcolour},   
    commentstyle=\color{codegreen},
    keywordstyle=\color{codepurple},
    numberstyle=\tiny\color{codegray},
    stringstyle=\color{codeorange},
    basicstyle=\ttfamily\footnotesize,
    breakatwhitespace=false,         
    breaklines=true,                 
    captionpos=b,                    
    keepspaces=true,                 
    numbers=left,                    
    numbersep=5pt,                  
    showspaces=false,                
    showstringspaces=false,
    showtabs=false,                  
    tabsize=4
}

\input{defs.txt}

\title[The SMART pulsar survey]{The Southern-sky MWA Rapid Two-metre (SMART) pulsar survey - II. 
Survey status, pulsar census, and  first pulsar discoveries}

\author{N. D. R. Bhat}
\affiliation{International Centre for Radio Astronomy Research, Curtin University, Bentley, WA 6102, Australia}
\author{N. A. Swainston}
\affiliation{International Centre for Radio Astronomy Research, Curtin University, Bentley, WA 6102, Australia}
\author{S. J. McSweeney}
\affiliation{International Centre for Radio Astronomy Research, Curtin University, Bentley, WA 6102, Australia}
\author{M. Xue}
\affiliation{National Astronomical Observatories, Chinese Academy of Sciences, Datun Road, Chaoyang District, Beijing 100101, China}
\author{B. W. Meyers}
\affiliation{International Centre for Radio Astronomy Research, Curtin University, Bentley, WA 6102, Australia}
\alsoaffiliation{Department of Physics \& Astronomy, University of British Columbia, 6224 Agricultural Road, Vancouver, BC V6T 1Z1, Canada}
\author{S. Kudale}
\affiliation{National Centre for Radio Astrophysics, Tata Institute of Fundamental Research, Pune 411 007, India}
\author{S. Dai}
\affiliation{Western Sydney University, Locked Bag 2751, Penrith South DC, NSW 1797, Australia}
\author{S. E. Tremblay}
\affiliation{National Radio Astronomy Observatory, 1003 Lopez Road, Socorro NM, 87801, USA}
\author{W. van Straten}
\affiliation{Institute for Radio Astronomy \& Space Research, Auckland University of Technology, Private Bag 92006, Auckland 1142, New Zealand}
\author{R. M. Shannon}
\affiliation{Centre for Astrophysics and Supercomputing, Swinburne University of Technology, P.O. Box 218, Hawthorn, VIC 3122, Australia}
\author{K. R. Smith}
\affiliation{International Centre for Radio Astronomy Research, Curtin University, Bentley, WA 6102, Australia}
\author{M. Sokolowski}
\affiliation{International Centre for Radio Astronomy Research, Curtin University, Bentley, WA 6102, Australia}
\author{S. M. Ord}
\affiliation{CSIRO Astronomy and Space Science, PO Box 76, Epping, NSW 1710, Australia}
\author{G. Sleap}
\affiliation{International Centre for Radio Astronomy Research, Curtin University, Bentley, WA 6102, Australia}
\author{A. Williams}
\affiliation{International Centre for Radio Astronomy Research, Curtin University, Bentley, WA 6102, Australia}
\author{P.~J.~Hancock}
\affiliation{Curtin Institute for Computation, Curtin University, GPO Box U1987, Perth, 6845, WA, Australia}
\author{R.~Lange}
\affiliation{Curtin Institute for Computation, Curtin University, GPO Box U1987, Perth, 6845, WA, Australia}
\author{J.~Tocknell}
\affiliation{AAO Macquarie, Macquarie University, NSW, Australia}
\author{M.~Johnston-Hollitt}
\affiliation{Curtin Institute for Computation, Curtin University, GPO Box U1987, Perth, 6845, WA, Australia}
\author{D.~L.~Kaplan} 
\affiliation{Department of Physics, University of Wisconsin--Milwaukee, WI 53201, USA}
\author{S. J. Tingay}
\affiliation{International Centre for Radio Astronomy Research, Curtin University, Bentley, WA 6102, Australia}
\author{M.~Walker}
\affiliation{International Centre for Radio Astronomy Research, Curtin University, Bentley, WA 6102, Australia}

\doi{XX.XXXX/pasa.XXXX.XX}

\received {dd Mmm YYYY}
\revised  {dd Mmm YYYY}
\accepted {dd Mmm YYYY}
\published{dd Mmm YYYY}

\begin{document}

%\begin{frontmatter}
%\maketitle

\input{02abstract}

%\end{frontmatter}

%\section{Introduction}
\input{02introduction}

%\section{Survey Status}
\input{02survey-status}

%\section{Confirmation and follow-up strategies }
\input{02confirmation-and-followup}

%\section{Pulsar discoveries and census  }
\input{02discoveries-and-census}

%\section{Summary and conclusions }
\input{02summary-and-conclusions}

%acknowledgements
\input{02acknowledgements}

%References 

%\bibliography{smart-pasa}
\bibliography{ms}

\input{figures}

%%%%%%%%%%%%%%%%%%%%%%%%%%%%%%%%%%%%%%%%%%%%%%%%%%%%%%%%%%%%%%%%%%%
\end{document}

%% file: defs.txt
\def\dmu{\ensuremath{\rm pc\,cm^{-3}}}
\def\lmu{\ensuremath{\rm mJy\,kpc^{2}}}
\def\rmu{\ensuremath{\rm rad\,m^{-2}}}
\def\TBhr{\ensuremath{\rm TB\,hr^{-1}}}

\def\fdotu{\ensuremath{\rm Hz\,s^{-1}}}  % spin freq. 1st deriv.
\def\pdotu{\ensuremath{\rm s\,s^{-1}}}  % spin period 1st deriv.
\def\sdlu{\ensuremath{\rm erg\,s^{-1}}} % spin-down luminosity
\def\us{\ensuremath{\mu{\rm s}}}  % microsecond
\def\ms{\ensuremath{\rm ms}}  % millisecond
\def\arcmin{\ensuremath{\prime}}
\def\arcsec{\ensuremath{{\prime\prime}}}
\def\mJy{\ensuremath{\rm mJy}}
\def\mJybm{\ensuremath{\rm mJy \, beam ^ {-1}}}
\def\uJybm{\ensuremath{\mu {\rm Jy \, beam} ^ {-1}}}
\def\deg{\ensuremath{{^{\circ}}}}

\def\sqdegphr{${\rm deg ^ {2} \, hr^{-1}}$\,}
\def\Tsys{\ensuremath{T_{\rm sys}}}

\def\Smin{\ensuremath{S _{\rm min}}}
\def\Smwa{\ensuremath{S _{\rm 150}}}

\def\Smeas{\ensuremath{S _{\rm 150} ^{\rm meas}}}
\def\Sexp{\ensuremath{S _{\rm 150} ^{\rm exp}}}
\newcommand{\psrone}{J0036$-$1033}
\newcommand{\psrtwo}{J0026$-$1955}
\newcommand{\psrthree}{J1002$-$2036}
\newcommand{\psrfour}{J1357$-$2530}

%% file: 02abstract
\begin{abstract}
In Paper I, we presented an overview of the Southern-sky MWA Rapid Two-metre (SMART) survey, including the survey design and search pipeline.  While the combination of MWA's large field-of-view and the voltage capture system brings a survey speed of $\sim$450\,\sqdegphr,  the progression of the survey relies on the availability of compact configuration of the Phase II array.
Over the past few years, by taking advantage of multiple windows of opportunity when the compact configuration was available, 
we have advanced the survey to 75\% of the planned sky coverage.
To date, about 10\% of the data collected thus far have been processed for a first-pass search, where 10 min of observation is processed for dispersion measures out to 250\,\dmu, to realise a shallow survey that is largely sensitive to long-period pulsars. The ongoing analysis has led to two new pulsar discoveries, as well as an independent discovery and a rediscovery 
of a previously incorrectly characterised pulsar, all from $\sim$3\% of the data for which candidate scrutiny is completed. 
In this sequel to Paper I, we describe the strategies  for further detailed follow-up including improved sky localisation and convergence to timing solution, 
and illustrate them using example pulsar discoveries. 
The processing has also led to re-detection of 120 pulsars in the SMART observing band, bringing the total number of pulsars detected to date with the MWA to 180, and these are used to assess the search sensitivity of current processing pipelines. 
The planned second-pass (deep survey) processing is expected to yield a three-fold increase in sensitivity for long-period pulsars, and a substantial improvement to millisecond pulsars by adopting optimal de-dispersion plans. 
The SMART survey will complement the highly successful Parkes High Time Resolution Universe survey at 1.2-1.5 GHz, and  inform future large survey efforts such as those planned  with the low-frequency Square Kilometre Array (SKA-Low). 
\end{abstract}

%% file: 02introduction
\section{INTRODUCTION }
\label{sec:intro}

%\textbf{
Large pulsar surveys have played a pivotal role in advancing pulsar astronomy. They have collectively led to a substantial increase in the population of known pulsars \citep[e.g.,][]{manchester1978,manchester2001,han2021}, with the discoveries of exotic objects (e.g., the double pulsar, the transitional millisecond pulsar) leading to transformational science \citep[e.g.,][]{lyne2004,kramer2021,archibald2009}. As such, conducting a full Galactic census and exploiting the specialised targets for high-profile science, such as testing strong-field gravity and detecting nanohertz-frequency gravitational waves, is a headline science theme for the Square Kilometre Array (SKA) and many of its precursor and pathfinder facilities \citep{keane2015,shao2015,janssen2015,meertime}. 
%}

The radically new designs of next-generation facilities bring significant changes (as well as challenges) to the pulsar survey landscape.  For instance, all-sky surveys, which have been mainly a single-dish arena for long, will now have to be conducted using distributed interferometric arrays or aperture arrays that make up these new generation facilities such as the Low Frequency Array (LOFAR; \citealt{lofar2013}) and the Murchison Widefield Array (MWA; \citealt{tingay2013,wayth2018}). They present a substantial complexity given the inherent  beamforming cost and large data rates, but also provide the benefits of significantly larger survey speeds, and new avenues for candidate confirmation and follow-up. At frequencies below 300 MHz, the ongoing LOFAR Tied-Array All-Sky (LOTAAS; \citealt{sanidas2019}) survey is the first such major step, while the South African MeerKAT telescope is proving to be highly successful in sensitive targeted searches \citep[e.g.,][]{ridolfi2022}. In the first paper of this series (hereafter referred to as Paper I), we described the Southern-sky MWA Rapid Two-metre (SMART) survey, a large project to search for pulsars and fast transients with the MWA's 140-170 MHz band, in the entire sky south of +30\deg in declination. 

While there has been substantial progress in terms of data collection effort, the SMART survey is currently in its early stages in terms of data processing and analysis. The adoption of voltage recording for survey data collection requires full-scale tied-array beamforming as the front-end of the processing chain, but it offers a number of unique benefits for candidate follow-up and confirmation. Most notably, the SMART survey has minimal reliance on new observations for immediate follow-up and/or confirmation of pulsar candidates. Access to the voltage data even allows accurate positional determination through a dense-grid localisation, using either survey data alone, or in combination with archival data, as vividly demonstrated in Paper I (see also \citealp{psrone}) through examples from early discoveries. 

Nevertheless, scientific merits of most new pulsar discoveries are typically revealed in due course of time, often through targeted follow-ups using more sensitive telescopes, or long-term timing observations. Two important pre-requisites are: 1) an accurate sky position, and 2) a full coherent timing solution. The latter typically relies on a series of regular observations carried out over the course of time. Again, with the advent of sensitive high-resolution imaging instruments such as the upgraded Giant Metre-wave Radio Telescope (uGMRT; \citealp{ugmrt}), and large survey data releases from sensitive imaging facilities such as the Australian SKA Pathfinder (ASKAP; \citealp[]{racs2020}), 
source identification and localisation can be expedited for arc-second level positional determination early on, thereby accelerating convergence to an initial timing solution. The design of the SMART also offers additional benefits, by virtue of its overlapping pointings, as detailed in Paper I. Moreover, when archival data are available, they can be utilised for an investigation of source variability, or for extending timing analysis, as demonstrated in \citet{psrone}.

%\textbf{
In Paper I, we described the survey design, science goals, current processing pipelines and follow-up strategies. 
In this sequel to Paper I, we present the survey status and further details on the initial discoveries, demonstrate and verify some of the detailed follow-up strategies, and provide a brief census of pulsar re-detections.
In \S~\ref{sec:surveystatus}, we present the survey status, and in \S~\ref{sec:confirmationandfollowup} details on confirmation and follow-up. Initial pulsar discoveries are summarised in \S~\ref{sec:discoveriesandcensus}, where we also present re-detections of a large sample of previously known pulsars from our SMART data processing. 
A summary is presented in \S~\ref{sec:summary}.  
%}

% FIGURE 1 %%%%%%%%%%%%%%%%%%%%%%%%%%%%%%%%%%%%%%%%%%%%%%%%%%%%%%%%
\begin{figure*}[t]
\begin{center}
%[width=\columnwidth]
\includegraphics[width=0.95\linewidth]%{mwa_obs_n70_res1_contour_pulsar_n125_pulsar_discovered_v2.png}
{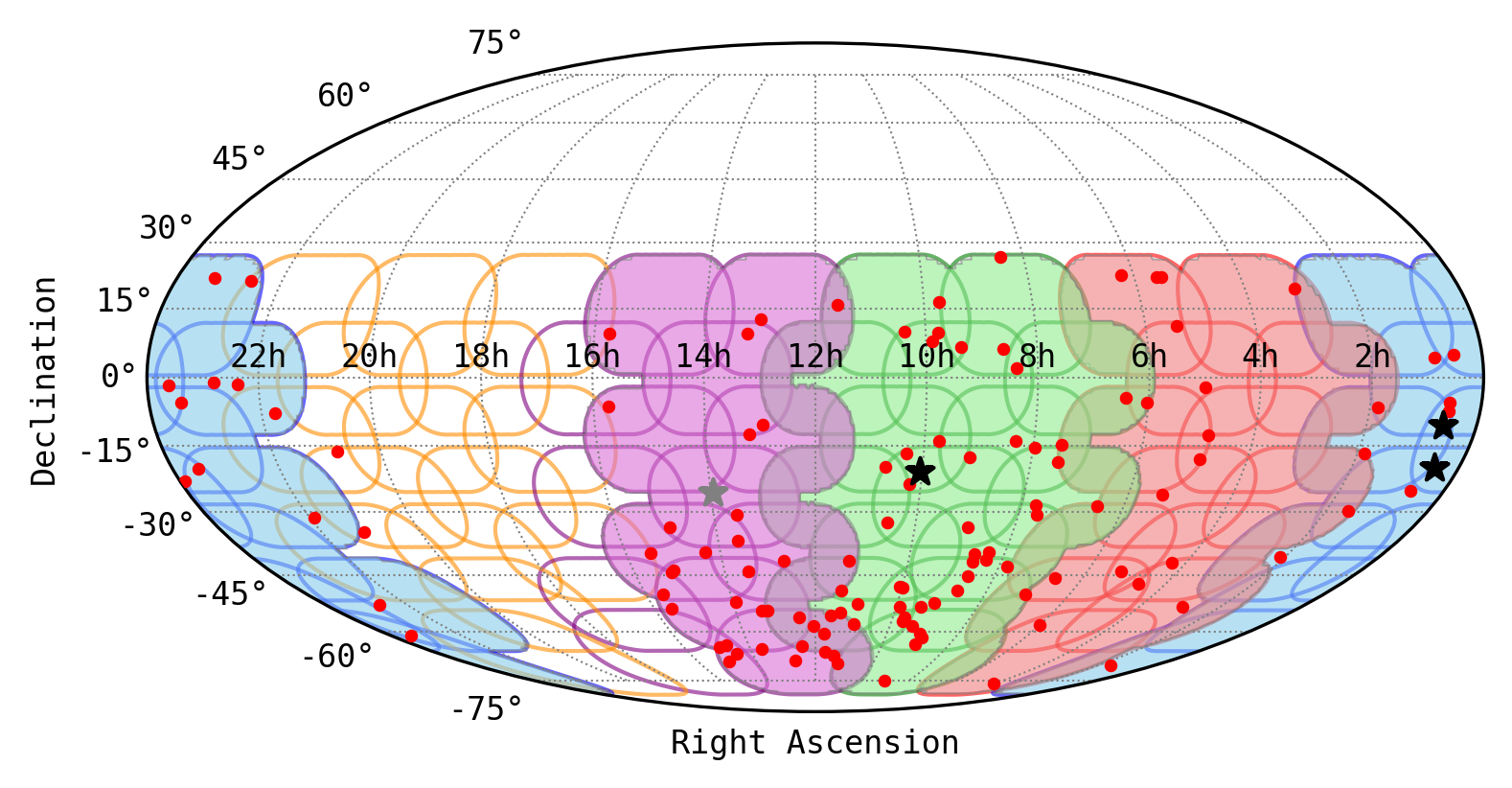}
\caption{
Sky plot summarising the observing strategy adopted for the SMART pulsar survey and the progress made to date: the full visible sky (i.e., declination $< +30\deg$ ) is covered using 70 pointings that overlap $10\deg$ in LST and $15\deg$ in declination. 
The coloured contours represent the half-power points of the main lobe of the tile (primary) beam (at 155\,MHz). 
The blue, red, and green pointing sets, as well as 10 of the 13 purple ones, have already been observed through four dedicated observing campaigns undertaken in the 2018B, 2019B, 2020A, 2021A semesters. 
The red filled circles are the known pulsar detections from the survey data (120 so far), and the black stars are the three new pulsar discoveries, including PSR~\psrtwo, an independent discovery from the SMART. The grey star is PSR~\psrfour, a re-discovery of 
a previously incorrectly characterised pulsar. 
%The grey filled circles are the known pulsar detections from the survey data (123 so far), and the red filled circles are the three new pulsar discoveries.
}
\label{fig:smartprogress}
\end{center}
\end{figure*}
% FIGURE 1 %%%%%%%%%%%%%%%%%%%%%%%%%%%%%%%%%%%%%%%%%%%%%%%%%%%%%%%%

%Table 2 of submitted version  
\input{table-smart-campaigns}

%% file: table-smart-campaigns
\begin{table*}[!t]
\caption{Summary of the survey observing campaigns to date} \label{tab:smartobs} \begin{tabular}{lcccccc}
\hline
% % % % Column names % % % %
  \multicolumn{1}{c}{Observing } &
  \multicolumn{1}{c}{Dates of observation} &
  \multicolumn{1}{c}{No. of VCS } &
  \multicolumn{1}{c}{RA range} &
  \multicolumn{1}{c}{ Data collected } & 
  \multicolumn{1}{c}{Completion } &
  \multicolumn{1}{l}{Redetected} 
\\
% % % % Column names % % % %
  \multicolumn{1}{c}{ semester} &
  \multicolumn{1}{c}{ } &
  \multicolumn{1}{c}{pointings} &
  \multicolumn{1}{c}{(hr)} &
  \multicolumn{1}{c}{ (TB) } & 
  \multicolumn{1}{c}{ status (\%)} &
  \multicolumn{1}{l}{pulsars$^{\dagger} $}  
\\
\hline  
2018B & September - December 2018 & 13 & 22-3 & 546 & 100 & 24 \\
2019B & September - November 2019 & 13 & 2-7 & 546 & 100 & 23(3) \\ 
2020A & January - March 2020 & 15 &  6-12 & 630 & 100 & 58(2) \\
2020B & April - May 2021 & 10 & 12-16 & 420 & 75 & 31(7) \\
\hline
\end{tabular}

\begin{tablenotes}
\item   $^{\dagger} $
The number in  parentheses are common redetections with the previous semester; e.g., 3 pulsars were detected in both 2018B and 2019B observations. 
\end{tablenotes}

\end{table*}

%% file: 02survey-status
\section{Survey Status }
\label{sec:surveystatus}

Even though the use of the voltage capture system (VCS) and the MWA's large field-of-view (FoV) means the data collection part of the survey can be completed in $<$100 hours of telescope time, practical considerations such as the data rate (42 TB per observation) and the availability of the compact configuration necessitate that the survey be carried out in multiple observing campaigns. By taking advantage of such windows of opportunity in the past few years, we have advanced the data collection effort to 75\%. Specifically, through four dedicated campaigns in the 2018B, 2019B, 2020A and 2021A semesters, we have covered much of the sky within the right ascension (RA) range $<$16\,hr and $>$21\,hr and declination south of $+30\deg$, as shown in Fig.~\ref{fig:smartprogress}. The fourth campaign to cover the 13-17\,hr RA range was only partially complete, as the system was no longer available for observations for an extended period since May 2021.  
Further details of the pointings and observing campaigns are summarised in Table~\ref{tab:smartobs}. 
The remaining 25\%  of the survey will be undertaken 
once the array returns to the compact configuration (by around mid-2023). 
Substantial progress has already been made with the science commissioning of the new corelator/voltage capture system (MWAX VCS), with the successful demonstration of the ability to attain full coherent beam sensitivity.

In the ongoing first-pass search, 10 min of data from each observation are processed in 2358 trial dispersion measures (DMs) out to 250\,\dmu, thus effectively performing a shallow survey of a large part of the southern sky. The limiting sensitivity achieved is comparable, or slightly better than, that of the Parkes 70cm survey from the 1990s \citep{70cm}, at least for pulsars with spectral index $\alpha \lesssim -1.6$. At the time of writing, 
processing is completed for $\sim$10\% of the data collected thus far, covering $\sim$55\% of the sky searched in the {\it shallow survey} phase, 
resulting in $\sim$1 million candidates, 
after filtering through the current machine-learning (ML) tools. Efforts are under way to substantially improve the efficacy of candidate filtering schemes, using semi-supervised generative adversarial network (SGAN) based classifiers \citep[e.g.,][]{sgan}, and will be integrated into the processing pipelines in the near future. Initial results are encouraging, and details including current implementation and the efficacy achieved, will be described in a future publication. 

\begin{figure*}[ht]
  \subfloat[]{\includegraphics[width=0.5\textwidth]
 {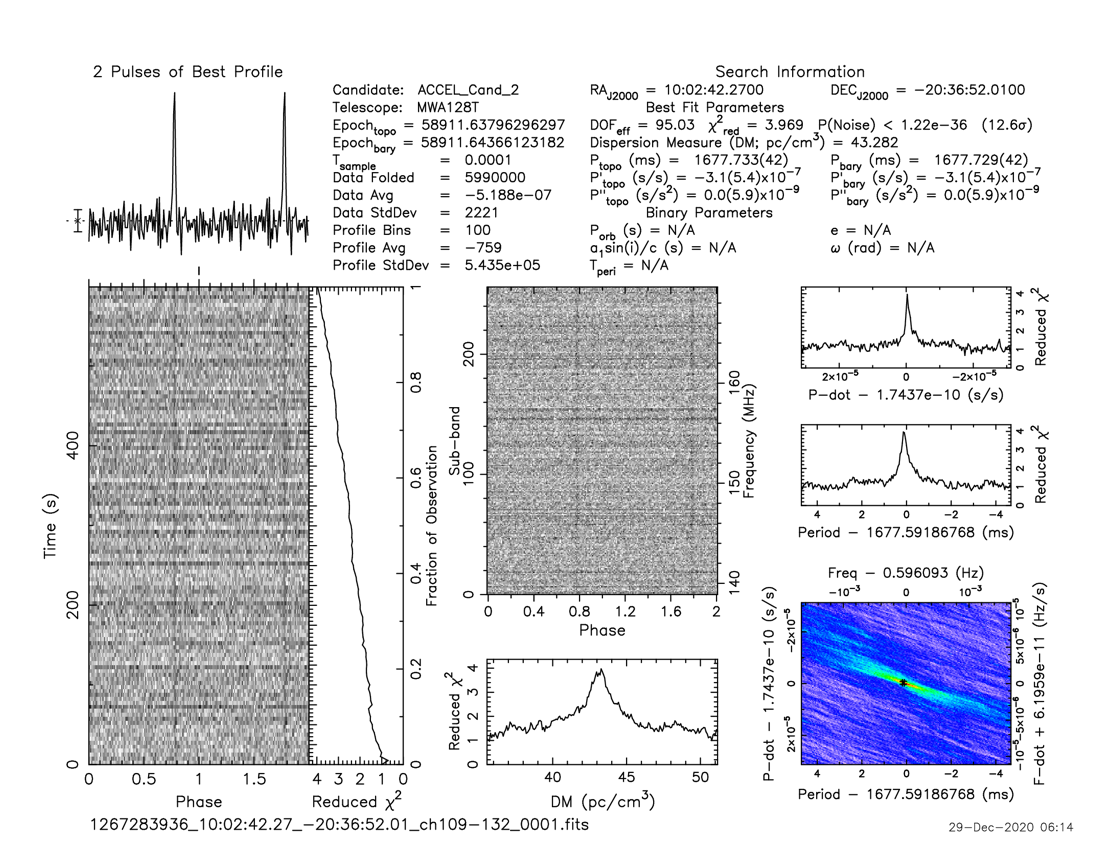}} 
 \hfill 	
  \subfloat[]{\includegraphics[width=0.5\textwidth]
 {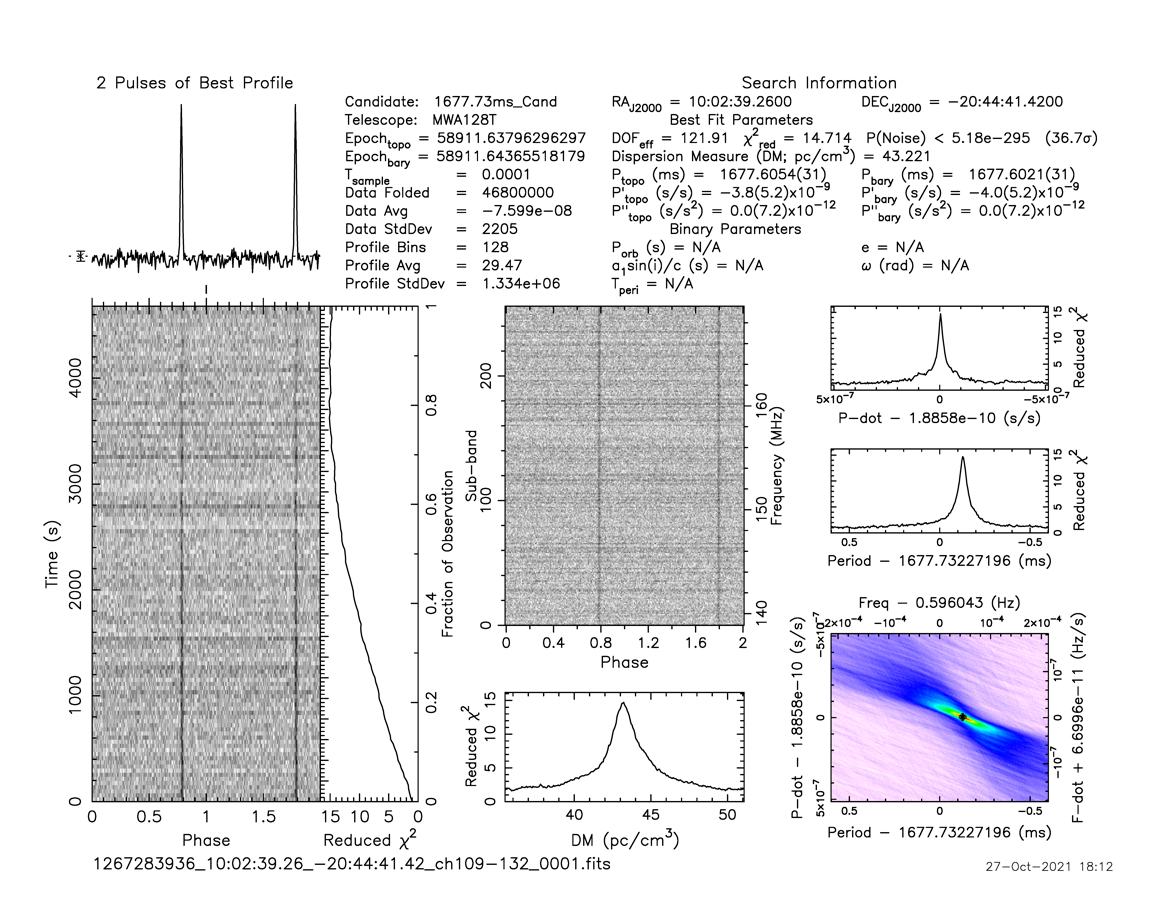}} 
 \newline
  \subfloat[]{\includegraphics[width=0.5\textwidth]
  {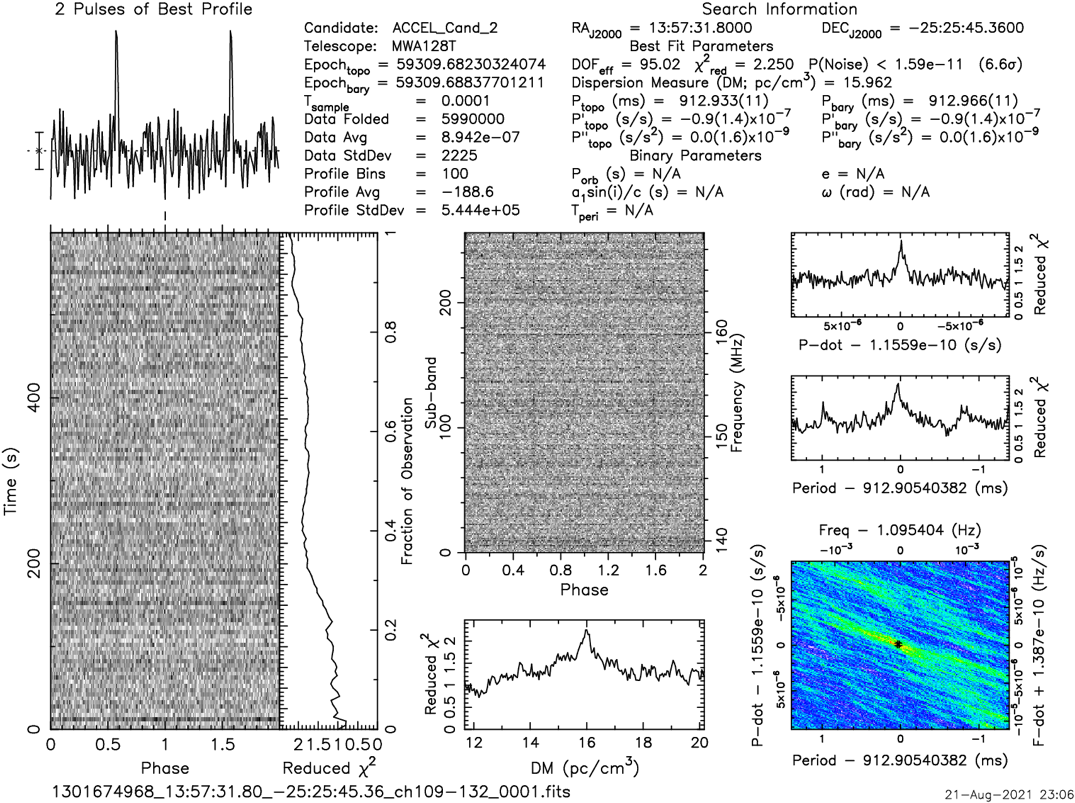}}
 \hfill	
  \subfloat[]{\includegraphics[width=0.5\textwidth]
{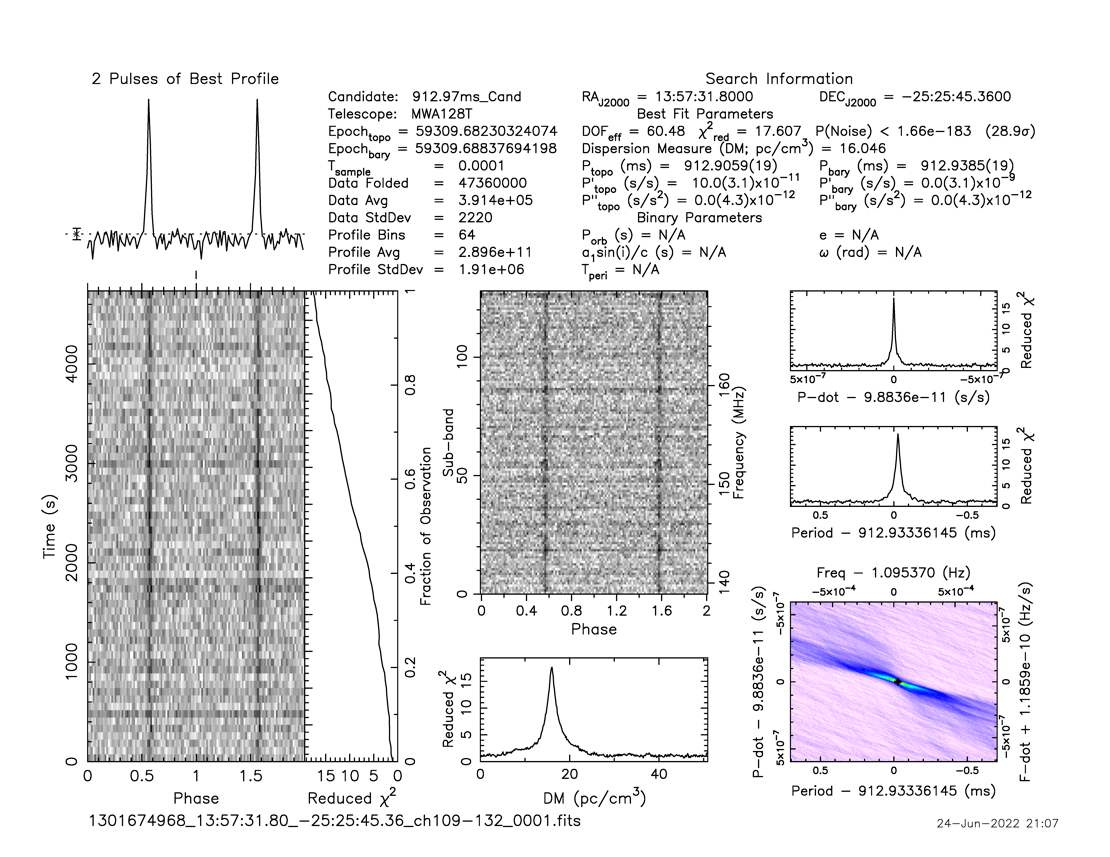}}
 \hfill 	
\caption{Detection plots of PSRs~\psrthree{} and \psrfour{} from SMART search processing. Panels (a) and (c) are diagnostic plots (from PRESTO) of the original candidate detections (PSR~\psrthree{} and PSR~\psrfour{}, respectively), while panels (b) and (d) show the improved detections from follow-up processing (confirmation) using the same inital detection observation, but by forming a tied-array beam for the full 80-min observing duration.}
\label{fig:prestoplots}
\end{figure*}

The current efficacy of ML tools means that a large volume of candidates require human scrutiny; however, only a small fraction ($\lesssim$3\%) of these have been examined for promising candidates to follow-up, resulting in four pulsar discoveries, including an independent discovery and a re-discovery of a previously incorrectly characterised pulsar, as well as re-detections of 120 previously known pulsars. 
All four discoveries have been followed up with Parkes and uGMRT, and also using archival/new data from the MWA, for early science extraction, and with the intent to develop and demonstrate effective follow-up strategies. Detection plots of PSRs~\psrthree{} and \psrfour{} are shown in Fig.~\ref{fig:prestoplots}. The parameters of the three new pulsars are summarised in Table~\ref{tab:smartnew}, and some of their detailed characteristics are described in the sections below.

\input{smart-new-pulsars}

%% file: smart-new-pulsars
%%%%%%%%%%%%%% Table 2 - timing solution %%%%%%%%%%%%%%%%%%%%%%%%%
\begin{table*}[!t]
\begin{threeparttable}
\caption{Parameter summaries of SMART pulsar discoveries} \label{tab:smartnew}
\begin{tabular}{lcccc}
\hline
% % % % Column names % % % %
  \multicolumn{1}{l}{Parameter} &
  \multicolumn{1}{c}{PSR~\psrone} &
  \multicolumn{1}{c}{PSR~\psrtwo$^{1}$} &
  \multicolumn{1}{c}{PSR~\psrthree} &
  \multicolumn{1}{c}{PSR~\psrfour$^{2}$}
\\
\hline  
Right ascension (J2000) & 00$^h$36$^m$15$^s$.01(4) & 00$^h$26$^m$36$^s$.3(2) & 10$^h$02$^m$39$^s$.3 & 13$^h$57$^m$24s \\
Declination (J2000) & $-$10$\deg$33$^{\arcmin}$14$^{\arcsec}$.2(9) & $-$19$\deg$55$^{\arcmin}$59$^{\arcsec}$.3(30) & $-$20$\deg$44$^{\arcmin}$41$^{\arcsec}$.4 &  $-$25$\deg$30$^{\arcmin}$39$^{\arcsec}$ \\
Galactic longitude ($l$) & 112.3\deg & 87.3\deg & 258.7\deg & 321.16\deg	\\
Galactic latitude ($b$) & $-$72.9\deg & $-$80.9\deg & 27.3\deg & 35.02\deg \\
Spin period ($P$) & 0.900009289(3)\,s & 1.30615183533(1)\,s & 1.677733\,s & 0.912\,s\\
Dispersion measure (DM) & 23.1(2)\,\dmu & 20.81(1)\,\dmu & 43.282\,\dmu & 16.046\,\dmu\\
Flux density at 400 MHz & 1\,\mJy & 4.2\,\mJy & -- & 0.5-1\,\mJy\\
Flux density at 1400 MHz & 0.1\,\mJy & 0.3\,\mJy & 0.1\,\mJy & -- \\
Spectral index ($\alpha$) & $-2.0\pm0.2$ & $-1.6$ & $\lesssim -2$ & --\\
Rotation measure (RM) & $-8.1\pm0.7$\,\rmu & $3.65\pm0.09$\,\rmu & -- & \\ 
\hline
\end{tabular}
\begin{tablenotes}
\item[$^{1}$] 
Independent rediscovery. The source was first detected as a GBNCC candidate \citep[cf.][]{mcsweeney2022}. 
A full timing solution from MWA data is presented in Table~\ref{tab:psr2ephem}.
\item[$^{2}$] 
Independent rediscovery of a previously incorrectly characterised pulsar. See text (\S~\ref{sec:pulsarfour}) for details.
\end{tablenotes} 
\end{threeparttable}
\end{table*}

%%%%%%%%%%%%%%%%%%%%%%%%%%%%%%%%%%%%%%%%%%%%%%%%%%%%%%%%%%%%%%%%%%%%%%

%% file: 02confirmation-and-followup
\section{Follow-up of pulsar discoveries } \label{sec:confirmationandfollowup}

%\textbf{
As described in Paper I, the SMART survey's unique design offers a range of flexible reprocessing avenues for candidate confirmation and improved positional determination, without having to rely on making immediate new observations. Furthermore, the archival data from past VCS projects can be leveraged for further detailed follow-up and characterisation, as demonstrated in \citet{psrone} and \citet{mcsweeney2022}; e.g., extending an initial timing solution derived from regular observations to determine the spin period ($P$) and its first time-derivative (${\dot P}$). An improved sky position is vital for follow-ups with more sensitive telescopes operating at higher frequencies, but a convergence toward this can be expedited via high-resolution imaging using facilities such as the uGMRT. 
%}

In the following sections we outline some additional strategies that are natural extensions to those already described in Paper I. These include: (1) an interferometric localisation via high-resolution imaging, (2) the processing of archival VCS observations for an extended timing analysis, and (3) high-frequency follow-up using sensitive telescopes such as the uGMRT and  Parkes. 
As we have demonstrated in earlier publications from the SMART survey, some of these are proving useful for rapid convergence to a working pulsar timing solution \citep[e.g.,][]{psrone}, and for ascertaining the detailed characteristics in terms of  scientific significance \citep[e.g.,][]{mcsweeney2022}.

% FIGURE 3 %%%%%%%%%%%%%%%%%%%%%%%%%%%%%%%%%%%%%%%%%%%%%%%%%%%%%%%%
\begin{figure*}[t]
\begin{center}
%[width=\columnwidth]
\includegraphics[width=0.45\linewidth]{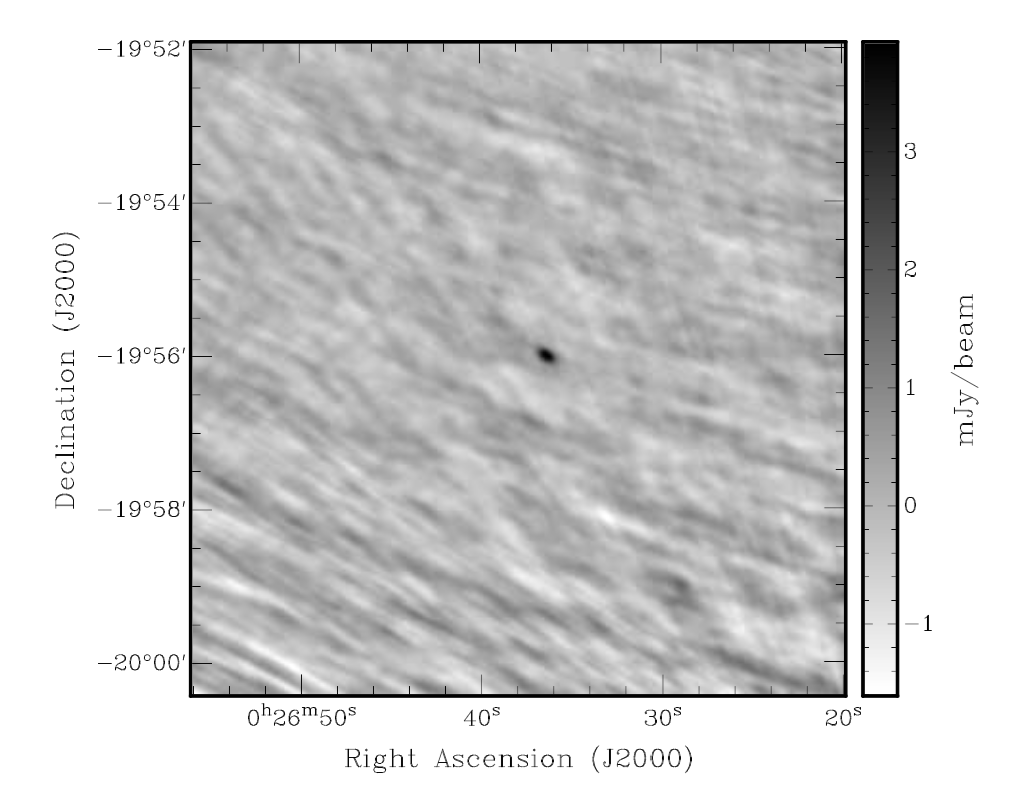}
\includegraphics[width=0.45\linewidth]{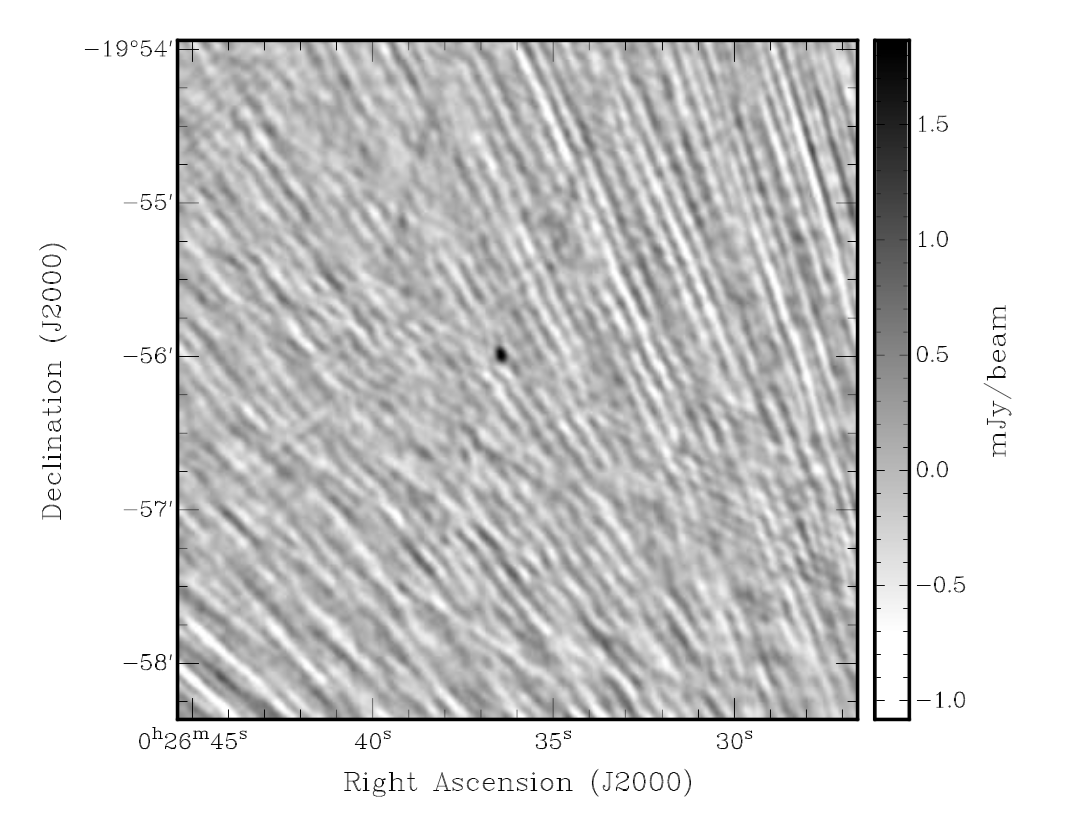}
\includegraphics[width=0.90\linewidth]{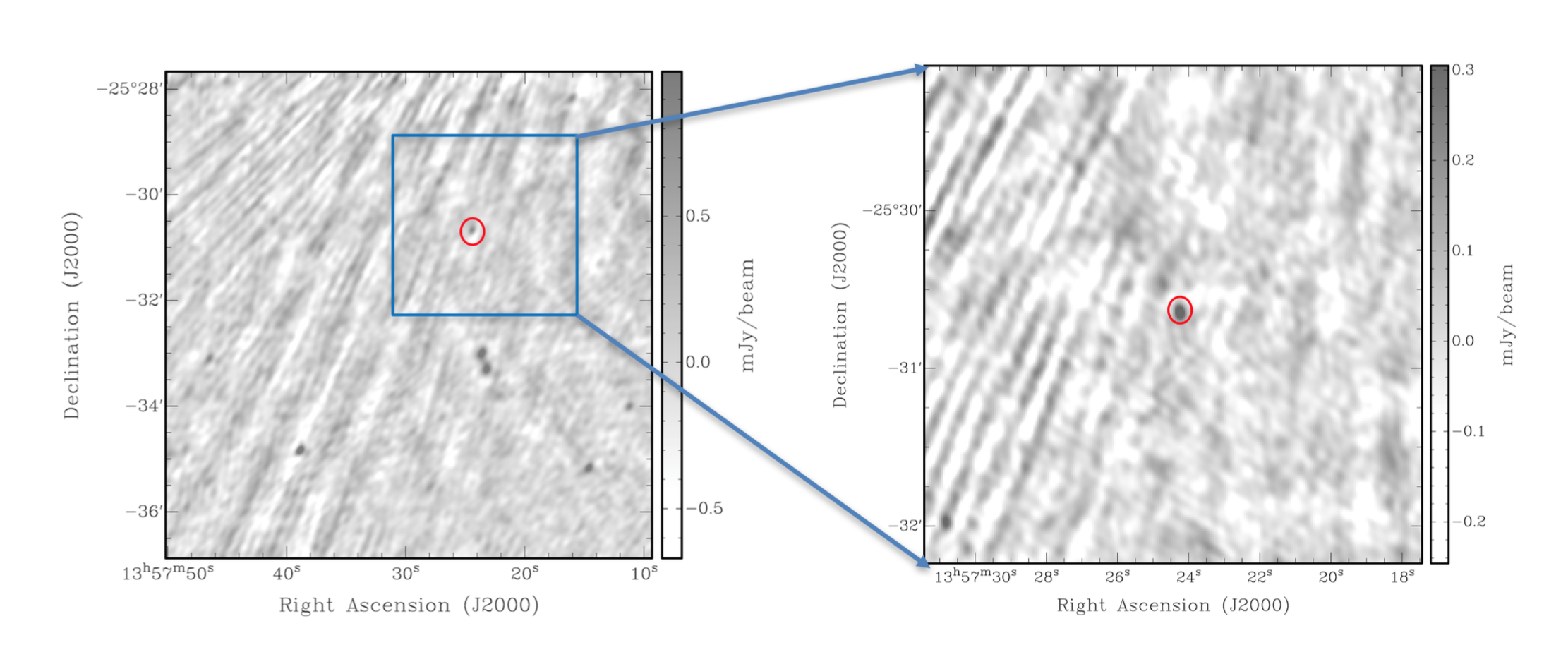}

\vspace{-0.6cm}
\caption{\emph{Upper panels}: uGMRT images of PSR~\psrtwo{} in Band 3 (300-500\,MHz; {\it left}) and Band 4 (550-750\,MHz; {\it right}), obtained from integration times of 136\,min and 56\,min, respectively. The rms sensitivity is $\approx$276\,\uJybm{} in Band 3, and $\approx$210\,\uJybm, respectively, yielding $\sim 15\sigma$ and $\sim 10\sigma$ detection of the pulsar. The presence of a bright nearby source (4\,Jy at 400\,MHz, at $\sim 10.5^\arcmin$ from the pulsar position) impacts the quality of calibration achievable, as seen in Band 4 observations.
Band 3 observations were made at an offset position of the pulsar, 
and the effect of the bright source was subdued owing to attenuation 
of the primary beam.
\emph{Lower panels}:
uGMRT images of PSR~\psrfour{} in Band 3 (300-500\,MHz; {\it left}) and Band 4 (550-750\,MHz; {\it right}), obtained from integration times of 90\,min and 71\,min, respectively. The rms sensitivity is $\approx$153\,\uJybm{} in Band 3, and $\approx$53\,\uJybm, respectively, yielding $\sim 6\sigma$ and $\sim 11\sigma$ detection of the pulsar. Band 3 imaging of the initial (poorly localised) pulsar position revealed multiple compact sources, one of which turned out to be the pulsar (see \S~\ref{sec:pulsarfour} for details).
}
\label{fig:psrtwoimage}
\end{center}
\end{figure*}
% FIGURE 3 %%%%%%%%%%%%%%%%%%%%%%%%%%%%%%%%%%%%%%%%%%%%%%%%%%%%%%%%

%\subsection{Imaging}
\subsection{Imaging for improved localisation}
\label{sec:imaging}

\input{table-imaging-summary-J0026}

\subsubsection{High resolution imaging with the uGMRT}
\label{sec:hires-imaging-gmrt}
Beyond the tied-array beam localisation as detailed in Paper I, 
the position of SMART-discovered pulsars can be further improved via high-resolution continuum imaging with interferometric arrays, for which the high sensitivity and angular resolution of the upgraded GMRT (uGMRT) is highly appealing, 
provided the source is north of $-55\deg$ in declination, and the spectrum is not ultra-steep (spectral index, $ \alpha \gtrsim -3 $). 
For PSR~\psrone, the first discovered MWA pulsar, this was convincingly demonstrated in \citet{psrone}, for which the localisation uncertainty was significantly improved from an initial $\sim 10^{\arcmin}$ (i.e., discovery observation) to $\sim 10^{\arcsec}$ through a dense-grid tied-array beam (TAB) localisation using the three different array configurations, and then eventually to $ \lesssim 1^{\arcsec}$ via high-resolution imaging of the pulsar using uGMRT Band 3 and 4 (300-750\,MHz) observations. 

For PSR~\psrtwo, even though uGMRT observations were made in similar bands using the identical setup (i.e., concurrent visibility recording and phased-array beams), and analysed using similar procedures, imaging of Band 3 data (300-500\,MHz) proved difficult due to the presence of a bright source 
($S_{400} \sim 4 $\,Jy) in the vicinity  ($\sim 10.5^{\arcmin}$ from the putative pulsar position). Fortunately, our earlier observations that were made as part of the DDTC185 project (and before the realisation that the original detection was in fact through one of the grating lobes), provided significant attenuation of the source owing to the initial $\sim 42 ^{\arcmin}$ (and hence incorrect) position. Analysis of these observations resulted in a $\sim 15\sigma$ detection, as shown in the left panel of Fig.~\ref{fig:psrtwoimage}, notwithstanding the fact that the pulsar was at more than half-power beam width offset from the phase centre. The resultant rms of $ 276\,\uJybm $ is nearly a factor of two larger than that was achieved with the imaging analysis of PSR~\psrone. 

The imaging of Band 4 data (550-750\,MHz) benefited from a much narrower primary beam (and hence a substantially higher attenuation toward the bright, nearby source), besides a reduced flux density due to spectral index, 
yielding a $\sim 10\sigma$ detection of the pulsar, as shown in the right panel of Fig.~\ref{fig:psrtwoimage}. The rms attained in this case is $\approx 210\,\uJybm $, almost an order of magnitude larger than that is typically attainable for deep imaging of 1\,hr in uGMRT Band 4. This illustrates the complexity in interferometric localisation, when the imaging analysis is hampered by bright sources in the vicinity.

A summary of uGMRT-derived positions are given in Table~\ref{tab:psrtwo}, along with the initial MWA-derived position. 
There is $\approx 32^{\arcsec}$  discrepancy between the uGMRT- and MWA-derived positions, which can be possibly attributed to a residual ionospheric refraction as discussed in \citet{swainston2022}. It is possible that the $\approx 3^\arcsec $ offset between the two uGMRT bands may also be caused by the ionosphere; even so, the final position, which we adopt as the mean of those from the two uGMRT bands, is still over an order of magnitude more precise than that reported by \citet{mcsweeney2022}. Further resolution to this will be possible once an improved (and fully coherent) timing solution becomes available.

\subsubsection{Imaging with the MWA}

For sources outside the uGMRT's sky ($\delta < -55\deg $), in principle, the recorded VCS data (if available) from long-baseline configurations of the array (Phase I or Phase II extended) can be correlated offline and imaged to verify the position obtained via the TAB localisation. However, as we demonstrated in \citet{psrone}, even with co-addition of multiple  observations we were unable to reach a sensitivity better than $\sim$6–8\,\mJybm{} (for Stokes I), and hence an imaging detection may be possible only for pulsars that are sufficiently bright (\Smwa$\,\gtrsim\,$30\,mJy). For long-period pulsars, such as PSR~\psrtwo, `on' and `off' gated imaging 
may also be attempted for improved detection in imaging, though this methodology is currently under development.

\subsection{Archival observations}
\label{sec:archival-observations}

Apart from the observations taken as part of the SMART campaign, the MWA archive includes many VCS observations dating back to 2014 when early pulsar and high-time resolution science observations commenced. 
These observations were typically parts of projects focusing on individual pulsars, so the distribution of the pointing directions across the sky is non-uniform, and some areas of the sky are not covered at all. Nevertheless, if a promising SMART candidate 
happens to be in an historically observed part of the sky,
then the archival VCS data can be used for further confirmation of the candidate without requiring extra observations be taken.

\begin{figure}
\centering 
\includegraphics[width=\columnwidth]{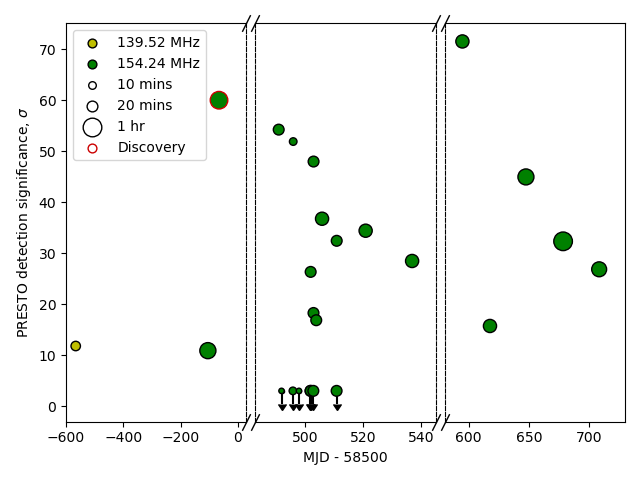}
\vspace{-0.6cm}
\caption{
Detection summary of PSR~\psrtwo, over a $\sim$3.5-year time span; the PRESTO detection significance ($\sigma$) is plotted against the MJD of observation. The detection are predominantly from observations over a 30.72\,MHz bandwidth centred at 154.24\,MHz, with time integration varying from 10\,min to 1\,hr. 
Note that the $\sigma$ reported for the discovery observation is for the full 80-minute data that was processed for follow-up and not for the 10-minute observation for initial (first-pass) search, in which the pulsar was discovered.
The 3-$\sigma$ upper limits indicate non-detection in a number of 10-20\,min observations.   
}
\label{fig:psrtwoarchival}
\end{figure}

Again we use PSR~\psrtwo{} as an example, whose location had fortuitously been covered by multiple previous observations. 
Six of these observations are included in the analysis described in \citet{mcsweeney2022}, but since then, more archival observations have been identified and processed, as shown in Fig. \ref{fig:psrtwoarchival}. 
Depending on the intervals between detections, their duration, and the quality of the detections themselves, such archival observations can be used to constrain otherwise hard-to-measure properties of the candidate pulsars which require long follow up campaigns, such as the flux variability due to scintillation, and their timing properties.

%{\bf
\subsubsection{Leveraging archival data to achieve timing coherence} \label{sec:timingexample}
As an example of using the archival MWA data and improved positions from imaging follow-up (see \S\ref{sec:imaging}), we are able to extract times-of-arrival (TOAs) from MWA observations alone and, in principle,  phase-connect a timing solution.
Here, we demonstrate that for PSR~\psrtwo{} across a $\sim$1.7-year time span.

Examining the MWA VCS observation database revealed observations containing the nominal pulsar position dating back to April 2017, including dedicated followup observations for PSR~\psrone{} in late-2020 into 2021.
These serendipitous observations are well suited to generating a phase-connected timing solution as there was a dense observing campaign where the same field was observed 8 times in three days (2020-10-26 to 2020-10-28), surrounded by regular observations spaced every 3-4 days for approximately 2 weeks before (2020-10-10 to 2020-10-22) and roughly weekly cadence for one month after (2020-10-30 to 2020-11-27).
Monthly cadence data were then collected until the the system became unavailable for science observing (see \S\ref{sec:surveystatus}).
These observing time scales are adequate to accurately determine the rotation frequency, detect spin-down and measure the DM.

The initial rotational characteristics presented by \citet{mcsweeney2022} are sufficient to fold PSR~\psrtwo{} for a single observation, but quickly lose phase connection after $\sim1$ day. 
Using the uGMRT imaging position (Fig.~\ref{fig:psrtwoimage}) and the initial estimates of the spin period and DM, we processed some of the more recent archival observations (two observations in April 2019) and the majority of the PSR~\psrone{} timing followup observations (17 observations spanning October-November 2020), creating 10-second subintegration archives with DSPSR.
These archives were then subsequently processed using PSRCHIVE routines to remove RFI, specifically using the {\tt paz} median-filtered bandpass algorithm.
Since PSR~\psrtwo{} is a prolific nuller \citep[cf.][]{mcsweeney2022}, we manually inspected each observation with {\tt pazi} and masked the null subintegrations to construct a high-quality average (integrated) profile. 
Out of the total 21 observations we processed for this work, 10 were excluded either due to non-detections (unsurprising for a pulsar with a >70\% nulling fraction) or very weak detections (S/N < 15 for the fully time- and frequency-averaged profile).
The five observations with the highest S/N were then combined and artificially phase aligned using {\tt psradd} and a basic frequency-averaged template was created using {\tt paas}, where two Gaussian components modelled the profile.

Each observation was then pre-processed such that there were four frequency subbands per epoch, from which TOAs were extracted using the {\tt pat} utility within PSRCHIVE.
The initial ephemeris used to fold the observations and the TOAs were passed to both the Tempo2 \citep{tempo2_1, tempo2_2} and PINT \citep{pint} pulsar timing packages.
With each timing package, we were able to phase-connect all available
observations to the archival data points from nearly two years earlier by updating the spin frequency, adding a spin frequency time-derivative ($28\sigma$ significance), and small adjustment to the DM.
Both packages produced consistent results and uncertainties.
Given the sparse sampling on longer time scales we did not fit for the pulsar sky position, however, as we accrue and/or process additional observations (with the MWA or other telescopes) we will be able to constrain the position to better than the current $\sim 3^\arcsec$ uncertainty obtained from the uGMRT imaging. 
The current best timing solution (from Tempo2) is presented in Table~\ref{tab:psr2ephem}.
%}
\input{smart-psr2-timing-solution}

\subsection{High-frequency follow-ups} 
\label{sec:highfreqfollowup}

Our ability to attain a positional accuracy of $\sim$1-2$^{\prime}$ from discovery observations, or even down to $\sim$10-20$^{\prime\prime}$ when archival VCS data are also available from the long-baseline configurations of the MWA (see Paper I),
facilitates high-frequency follow-ups with sensitive telescopes such as the uGMRT and Parkes that share substantial common skies with the MWA. The combination of these telescopes and their wide-band instrumentation allow nearly seamless coverage in frequency from $\sim$100\,MHz to 4\,GHz, thereby enabling a detailed characterisation of spectral and propagation properties of new pulsar discoveries. For instance, in the case of PSR~\psrone, such follow-up observations helped establish the steep-spectrum 
($ S _{\nu} \propto \nu^{-2.0 \pm 0.2} $, where $ S _{\nu} $ is the flux density at frequency $ \nu $)
low-luminosity nature of the object (with a pseduo-luminosity $ L_{1400} \sim 0.1 $\,\lmu{} at 1400\,MHz). Observational studies of this kind are an integral part of our follow-up strategies, and are important to our understanding of the Galactic pulsar population in general.

\begin{figure*}[tp]
\begin{center}
\includegraphics[width=\linewidth]{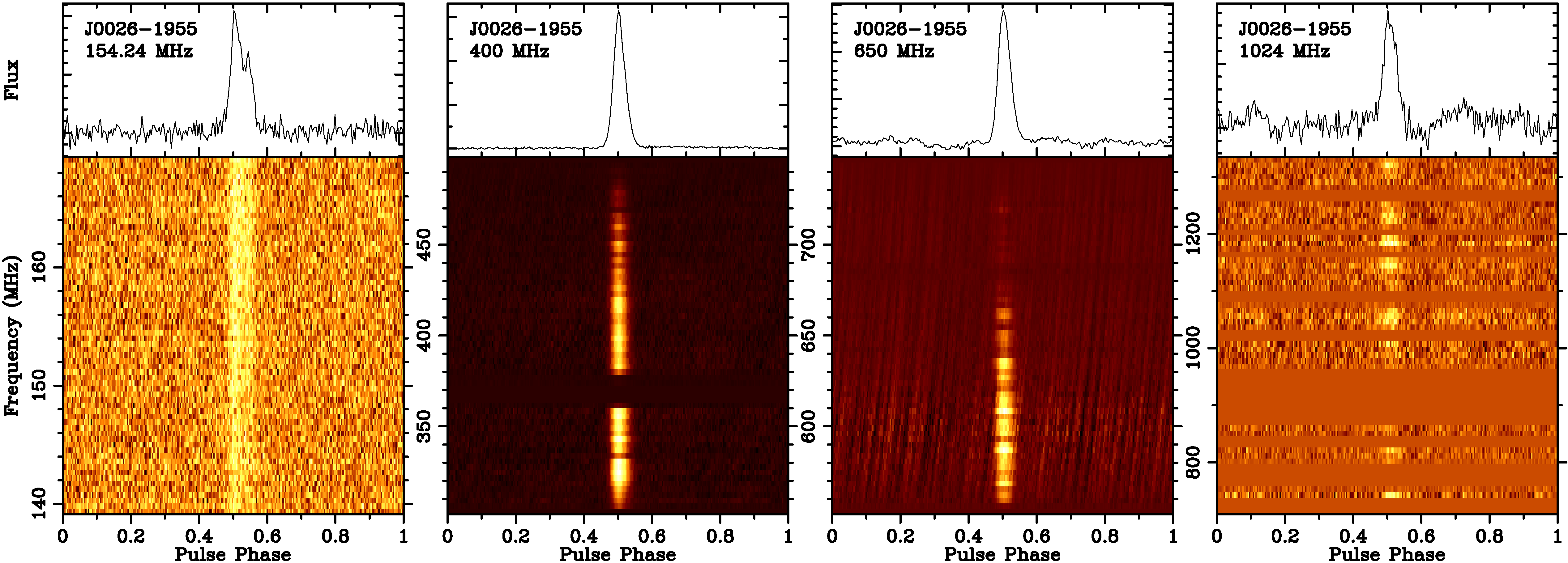}
\vspace{-0.7cm}
\caption{
Detection plots of PSR~\psrtwo{} with the MWA (left panel), uGMRT (two central panels), and Parkes (right panel) telescopes, spanning a frequency range from 140 to 1400\,MHz: the top panel is the integrated pulse profile and the waterfall plot below shows the pulse strength vs. pulse phase and frequency.  MWA observations were made with the Phase 2 compact configuration of the array, whereas those with the uGMRT made use of the 200\,MHz mode of the phased-array beamformer comprised of 11-13 antennas located within the central square. Parkes observations were made using the Ultra-Wideband Low-frequency receiver (700-4032\,MHz); however, the pulsar was detected only in the band below 1.4\,GHz. The profiles each have been downsampled to 256 bins across pulse phase and 64 frequency channels for plotting clarity.
}
\label{fig:psrtwo-allfour}
\end{center}
\end{figure*}

Example detections  from such high-frequency follow-ups are shown in Fig.~\ref{fig:psrtwo-allfour} for PSR~\psrtwo. These were obtained using  data collection and analyses procedures similar to those described in \citet{psrone}, i.e., concurrent imaging and phased-array observations using the uGMRT Band 3 (300-500\,MHz) and Band 4 (550-750\,MHz) receivers, and observations using the Parkes (\emph{Murriyang}) telescope. Observations at higher frequencies were made in the search mode of the Parkes ultra-wideband low-frequency (UWL; \citealp{hobbs2020}) receiver that spans the frequency range from 704 to 4032\,MHz, and followed analysis procedures as described in \citet{dai2019}. 

% FIGURE psrtwo all four band4  %%%%%%%%%%%%%%%%%%%%%%%%%%%%%%%%%%%%%
% FIGURE %%%%%%%%%%%%%%%%%%%%%%%%%%%%%%%%%%%%%%%%%%%%%%%%%%%%%%%%%%%

The best detection was made in the uGMRT Band 3 frequency range, where S/N$\sim$300 was achieved in one of the observing sessions in $\sim$30-min integration. Despite the severe RFI contamination of the Parkes data, the pulsar was clearly detected in the low band range 750-1300\,MHz, as shown in Fig.~\ref{fig:psrtwo-allfour}. Estimated flux densities are $\sim$4.2 and 2\,\mJy, respectively at 400 and 650\,MHz (from uGMRT imaging, see Fig.~\ref{fig:psrtwoimage}). The pulsar was also detected in the TIFR GMRT Sky Survey \citep{tgss2017}, with a flux density of $\sim$19\,\mJy{} (at 150\,MHz) and also in the ASKAP-low and -high bands, with flux densities of $\sim$1.3\,\mJy{} (at 888\,MHz) and $\sim$0.3\,\mJy{} (at 1.5\,GHz), respectively (E.~Lenc, private communication). The combination of these measurements suggest a spectral index $\alpha \sim -1.6$ for this pulsar, assuming no turnover down to $\sim$150\,MHz. However, as reported in \citet{mcsweeney2022}, the pulsar exhibits a large nulling fraction of $\sim$75\%, and complex sub-pulse drifting  behaviour, including rapid changes of the drift rate and consistent evolution within its two main drift modes. Further analysis using these data are deferred to a future publication (Janagal et al. in prep.).

%% file: table-imaging-summary-J0026
\begin{table*}[!t]
\begin{threeparttable}
\caption{Localisation summary of PSR~\psrtwo} \label{tab:psrtwo}
\begin{tabular}{lccc}
\hline
% % % % Column names % % % %
  \multicolumn{1}{l}{Telescope/System} &
  \multicolumn{1}{c}{Method} &
  \multicolumn{1}{c}{Right Ascension (J2000)} &
  \multicolumn{1}{c}{Declination (J2000)} 
\\
\hline  
MWA VCS & Tied-array beams & $00^h26^m37.5^s$ & $-19\deg 56^\arcmin 24.9^\arcsec$  \\
uGMRT Band 3  & Imaging & $00^h26^m36.2^s$ & $-19\deg 55^\arcmin 59.6^\arcsec$ \\
uGMRT Band 4  & Imaging & $00^h26^m36.41^s$ & $-19\deg 55^\arcmin 59.0^\arcsec$ \\
Final position & Imaging & 00$^h$26$^m$36$^s$.3(2) & $-$19$\deg$55$^{\arcmin}$59$^{\arcsec}$.3(30) \\ 
\hline
\end{tabular}
\end{threeparttable}
\end{table*}

%%%%%%%%%%%%%%%%%%%%%%%%%%%%%%%%%%%%%%%%%%%%%%%%%%%%%%%%%%%%%%%%%%%%%%

%% file: smart-psr2-timing-solution
\begin{table*}
\begin{threeparttable}
\caption{Timing solution for PSR \psrtwo{} using Tempo2.} \label{tab:psr2ephem} 
\begin{tabular}{lc}
\hline
\multicolumn{2}{c}{Timing solution and best-fit metrics} \\
\hline\hline
J2000 Right ascension (hh:mm:ss)& 00:26:36.3(2)$^\dagger$ \\  % 3 arcsec ~ 0.2 secs in HA
J2000 Declination (dd:mm:ss) & $-$19:55:59.3(30)$^\dagger$ \\
Spin frequency, $f$ ($\rm Hz$) & 0.765607774650(8) \\
First spin frequency time-derivative, $\dot{f}$ ($\times 10^{-16}$\,\fdotu) & $-$2.58(9) \\
Dispersion measure, DM (\dmu) & 20.81(1) \\
Reference epoch (MJD) & 58724 \\
Observing timespan (MJD) & 58427-59020 \\
Solar system ephemeris  & DE430 \\
Time ephemeris & FB90 \\
Clock standard  & TT(BIPM) \\
Units & TDB \\
Reduced-$\chi^2$ & 1.32 \\
Degrees of freedom & 40 \\
RMS timing residual (\us) & 1332 \\
\hline
\multicolumn{2}{c}{Derived quantities}\\
\hline
Galactic longitude, $l$ (\deg) & 83.2963 \\
Galactic latitude, $b$ (\deg)  & $-$80.8295 \\
Spin period, $P$ ($\rm s$) & 1.30615183533(1) \\
First spin period derivative, $\dot{P}$ ($\times 10^{-16}$\,\pdotu) & 4.4(1) \\
Characteristic age, $\tau_{\rm c}$ ($\rm Myr$) & 47 \\
Surface magnetic field strength, $B_{\rm S}$ ($\rm \times 10^{11}\,G$) & 7.7 \\
Spin-down luminosity, $\dot{E}$ ($\times 10^{31}$\,\sdlu) & 2.3 \\
\hline
\end{tabular}
\begin{tablenotes}
\item[$^\dagger$] The position of the pulsar was fixed at the uGMRT imaging position, as given in Section~\ref{sec:imaging}.
\end{tablenotes}
\end{threeparttable}
\end{table*}

%% file: 02discoveries-and-census
\section{Pulsar discoveries and census  }
\label{sec:discoveriesandcensus}

As described in \S~\ref{sec:surveystatus}, a first-pass search processing is currently in progress, to realise a shallow survey for long-period pulsars. About $\sim$10\% of the data
collected so far have been processed in this mode, though it covers $\sim$55\% of the planned sky coverage.  Only a small fraction ($\lesssim$3\%) of the candidates have been examined for promising candidates to follow up, resulting in four pulsar discoveries, one of which is an independent discovery of a GBNCC pulsar candidate, and another a re-discovery of a pulsar that was originally found in the Parkes 70cm survey \citep{70cm} but reported  with an incorrect name and DM. 
The processing has also led to a re-detection of 120 previously known pulsars, which brings the number of MWA pulsar detections to 180, a significant improvement over the initial census (at 185 MHz) reported in \citet{xue2017}. 
All four pulsars are being followed up with Parkes and the uGMRT, and also using archival and new data from the MWA, for detailed characterisation including their timing, variability,
and early science, and also with the intent to develop and demonstrate effective follow-up strategies. 
Basic parameters of these pulsars are summarised in Table~\ref{tab:smartnew}, and we elaborate on some specifics in the sections below.

\subsection{Individual pulsar details}

\subsubsection{PSR~\psrone}
\label{sec:pulsarone}
PSR~\psrone{} is the first new pulsar discovery from the SMART survey, the details and initial follow-up of which have been reported in \citet{psrone}. This non-recycled  pulsar with a period $P=0.9$\,s and a ${\rm DM}=23.123\,\dmu$ has a fairly typical magnetic field strength ($ B \sim 4.4 \times 10^{11}$\,G) 
and characteristic age ($\tau _c \sim 67$\,Myr) but has a spectral 
index $\alpha = -2.0 \pm 0.2 $. An estimated luminosity at 1400\,MHz, $L_{1400}$ $\sim$0.1\,\lmu,  places this pulsar in the lowermost two percentile of the currently known population of long-period pulsars in luminosity. Its DM and a location at a high Galactic latitude ($b \approx -72.9\deg$) make it an excellent target for scintillation and variability studies for probing the nature of plasma turbulence in a region where it is poorly constrained. The related observations are currently under way at the uGMRT and the Five-hundred-metre Aperture Spherical Telescope (FAST) and will be reported in a future publication. 

\subsubsection{PSR~\psrtwo}
\label{sec:pulsartwo}
The ``discovery'' of PSR~\psrtwo{} revealed some remarkable features of the SMART survey; most notably, the complexity of the tied-array beam pattern, resulting from a large fraction of the tiles configured as hexagonal layouts (see Paper I; Figs. 7 and 8).
This can result in multiple detections of pulsars through the near and far side (grating) lobes, especially when the pulsar is sufficiently bright.  For instance, PSR~\psrtwo, which was initially detected as a 11$\sigma$ candidate (in a 10-min observation) was eventually re-detected in a different pointing, almost $\sim$45$^{\prime}$ offset from the initial position, and with a much higher significance of $\rm S/N = 60$,  leading to the revelation that the initial detection was in fact through one of the grating lobes. 

The pulsar was also particularly noted for its sub-pulse drifting nature which became the focus of immediate investigation. Incidentally, this turned out to be an independent discovery, as the pulsar was first reported as a candidate in the Green Bank Northern Celestial Cap (GBNCC) survey data; nevertheless, it was blindly discovered in our SMART processing. Further details of the pulsar, including analysis of its intriguing sub-pulse drifting and long-duration nulls, are reported in \citet{mcsweeney2022}.

\subsubsection{PSR~\psrthree}
\label{sec:psrthree} 
The discovery of PSR~\psrthree{} followed the new (revised) scheme for candidate selection and scrutiny, as described in Section 3.3 of Paper I. 
It is a long-period pulsar with $P=1.67$\,s and a moderate $\rm DM = 43.172$\,\dmu. The NE2001 model of electron density \citep{ne2001} places it at a distance of $\sim$1.4\,kpc. Its non-detection in the Parkes 70\,cm survey implies a spectral index $\alpha \lesssim -1.8 $, and non-detections in the TGSS (TIFR GMRT Sky Survey) and NVSS (NRAO VLA Sky Survey) continuum imaging sky surveys, which reached rms sensitivity limits of $\sim$5\,\mJybm{} and $\sim$0.5\,\mJybm{} (in the pulsar vicinity) at their respective frequencies (i.e. 150 MHz and 1400 MHz), implying a flux density at 150\,MHz, $ S_{150} \lesssim 10 $\,\mJy. This hints at the possibility of the pulsar being another low-luminosity object, with an implied 
$ L_{1400} \lesssim 0.3 $\,\lmu, which is confirmed by our recent Parkes detection, suggesting $ S_{1400} \sim 0.1$\,mJy, and hence akin to PSR~\psrone. The pulsar is currently being followed up for timing and localisation (via imaging) using Parkes and the uGMRT.

\subsubsection{PSR~\psrfour{}}
\label{sec:pulsarfour} 

\input{table-imaging-summary-J1357}

Of the 120 redetections from SMART data processing (see \S~\ref{sec:redetections}), PSR~\psrfour{} warrants some additional contextual discussion. It was first ``discovered'' in the untargetted SMART search workflow as a 6$\sigma$ candidate in a 10-minute observation, with a period ($P$) of 0.912\,s and DM of 16.046\,\dmu (see Fig.~\ref{fig:prestoplots}). Follow-up processing of the full 80-min observation resulted in a significantly improved detection, with a four-fold increase in S/N, as shown in the figure. This pulsar  has a small duty cycle of $\sim$2-3\%, and the DM implies a distance of $\sim$0.8\,kpc as per the NE2001 model of \citep{ne2001}.  However, a closer examination of the published literature including those from  low-frequency surveys with Parkes and the Green Bank Telescope (GBT), has essentially confirmed that this is a rediscovery. The Parkes 70\,cm survey appears to have been reported this as PSR~J1358$-$2533 with a similar period but with a  significantly discrepant DM = $28\pm3$\,\dmu\ and at $\sim$20$^{\prime}$ offset from our MWA position \citep{70cm}. The DM was subsequently revised upward by \citet{70cm_3} to $31.3 \pm 1$\,\dmu.  It appears that GBNCC has re-found the pulsar but with  a $\rm DM = 16.0 \pm 0.2\,\dmu$ \citep{McEwen2020}.  Even so, the position (and hence the pulsar's exact name) remained uncertain and a timing solution is currently not available. The pulsar also appears to be another low-luminosity object, even if we are to assume a spectral index ($\alpha$) of $-1.6$.

%new text following Cycle 43 analysis and results 
%{\bf 
In order to resolve this ambiguity, we 
recently undertook uGMRT follow-up observations of this pulsar (GTAC Cycle 43), where we used an observing set-up that is similar to those described in \S~\ref{sec:imaging} (i.e. concurrent imaging and phased-array beams). Following an initial non-detection, which is attributable to a poorly localised position, imaging of the recorded visibility data revealed multiple (five) compact sources in the field within $\sim 5 ^{\prime}$ from the phase centre; one of these turned out to be the pulsar, as shown in the lower panels of Fig.~\ref{fig:psrtwoimage}. This was confirmed in subsequent observations, where we cycled through the prospective sources by making separate phased-array beam observations.
Table~\ref{tab:psrfour} gives a summary of the pulsar positions from our uGMRT imaging along with the initial MWA position (from the TAB localisation). The pulsar position, with R.A. 13h57m24s, and decl. $-25\deg30^{\prime}39^{\prime\prime}$, while precisely determined down to $\lesssim 0.3^{\prime\prime}$, has a systematic offset of $\sim$3$^{\prime\prime}$, possibly due to ionospheric refraction (cf. \citealt{swainston2022}). The estimated flux densities are $\sim$1\,mJy and $\sim$0.5\,mJy at 400 and 650\,MHz, respectively.   
A summary of the pulsar parameters
are included in Table~\ref{tab:smartnew}. The new position  warrants renaming the pulsar to PSR~\psrfour. 
%}

\subsection{Redetections of known pulsars}
\label{sec:redetections} 

As described earlier, a secondary goal of the SMART survey is to map out the southern sky for low-frequency detections of known pulsars. 
This is important for two main reasons. Firstly, all past large pulsar surveys in the southern hemisphere have been at frequencies $\gtrsim$400\,MHz \citep[e.g.,][]{manchester1978,70cm,manchester2001,keith2010}. With the majority of the pulsars in the southern sky originally found and followed up at such high frequencies, low-frequency information ($\lesssim $300\,MHz) is lacking for a substantial number of known pulsars, especially for those at declination, $\delta \lesssim -30\deg $. Despite the MWA opening  a new low-frequency window for pulsar astronomy in the south, the limitations of legacy VCS (e.g., a maximum recording capacity of 1.5\,hr)
\footnote{The new high-time-resolution system (MWAX VCS) is capable of recording over much longer durations, up to $\sim$5-6 hours.}
and the logistics of handling large data rates prevented undertaking a complete pulsar census in an efficient manner. In an effort to partly address this, and as a preparatory work toward SKA-Low science, \citet{xue2017} carried out the first pulsar census with the MWA at 185\,MHz, using the then available resources and data sets (covering $\sim$55\% of the sky) and software and processing capabilities (i.e., incoherent detection, reaching $\lesssim$10\% of the full coherent beam sensitivity). This led to the low-frequency detections of 50 pulsars (including 6 millisecond pulsars), 10 of which were detected for the first time at low frequencies. This modest sample was also used for simulation studies and forecasting the detectable population of pulsars with SKA-Low, as summarised by \citet{xue2017}.  

\begin{figure*}
\centering
%\includegraphics[width=\linewidth]{mwa_pulsardetections_allskyplot.png}
%\includegraphics[width=\linewidth]{mwa_obs_n0_res3_minlines_pulsar_n186_pulsar_discovered.png}
%\includegraphics[width=\linewidth]{mwa_obs_n0_res3_minlines_pulsar_n124_pulsar_discovered.png}
%following BM comments
\includegraphics[width=\linewidth]%{mwa_obs_n0_res3_minlines_pulsar_n22_pulsar_discovered.png}
{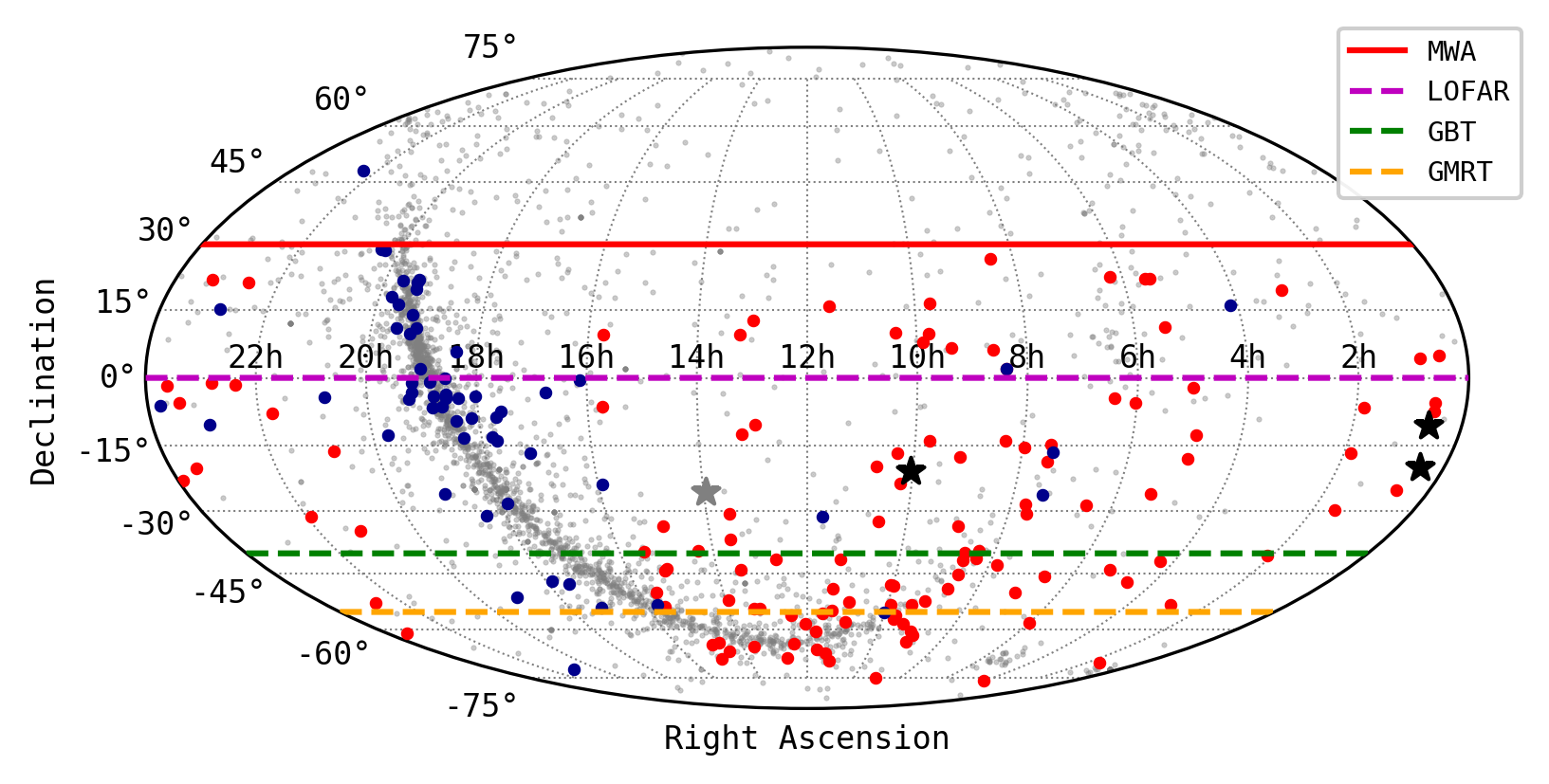}
\caption{
All-sky distribution of known pulsars in the ATNF pulsar catalogue (grey filled circles). The 120 pulsars detected in the SMART processing are shown as red filled circles, along with 57 pulsars detected from the processing of non-SMART MWA observations (shown as filled circles in dark blue, respectively, for incoherent and coherent beam detections).  The black star symbols are the three new pulsar discoveries from the SMART,
and the grey star is re-discovery of a previously incorrectly characterised pulsar. 
The declination limit of the SMART survey ($\delta < +30\deg$) is shown as the solid red line and for surveys with other (northern) telescopes, they are shown as dashed lines at the respective lower limits in declination (i.e., $\delta > 0\deg$ for LOTAAS, $\delta > -40\deg$ for GBNCC and $ -40\deg > \delta > -55\deg$ for GHRSS). 
}
\label{fig:mwapulsars}
\end{figure*}

A summary of re-detections from the SMART processing thus far is shown in Fig.~\ref{fig:mwapulsars}, where we have also shown those from non-SMART
observations (i.e., archival data or other VCS-based projects). In total there have been 120 re-detections from the SMART project; integrated profiles of these (at 155\,MHz) are shown in Fig.~\ref{fig:smartprofspage1}. Table~\ref{tab:SMARTknownPSR} provides a summary of these detections, along with the pulsar parameters $P$ and ${\rm DM}$, and the signal-to-noise ratio of detection $\rm (S/N)_{det}$. Given the strong direction and elevation dependencies of the MWA's sensitivity, we have accounted for the expected degradation in $\rm (S/N)_{det}$ due to the pulsar's offset within the primary beam (relative to the beam centre), while estimating the mean flux density at the observing frequency (listed in the column 6 of Table~\ref{tab:SMARTknownPSR}). 
As a result of the large extent of the MWA's primary beam, these offsets can be as large as $\sim$10-20\deg{} relative to the centre of the beam; for example, PSR~J0034$-$0721, a bright long-period pulsar (and a well-known sub-pulse drifter; \citealt{mcsweeney2017}), which would be detected with $\rm S/N\sim 400$ in $\sim$1\,hr observation, will likely be detected with 
$\rm (S/N)_{det} \lesssim 100$ when observations are made at an elevation of $\sim$30\deg{} 
and at $\sim$10\deg{} offset from the beam centre.

% FIGURE smart + non-smart pulsars %%%%%%%%%%%%%%%%%%%%%%%%%%%%%%

Including 57 pulsars from non-SMART observations, 18 of which are from targeted observations toward PSR~J1822-2256 (from the G0071 project; by Janagal et al.), this brings the current tally to 180 pulsars detected by the MWA. This is more than triple the number of pulsar detections with the MWA since the publication of \citet{xue2017}, but with the important distinction that many of these are full coherent beam detections, and the data can be reprocessed for full polarimetric information and/or higher time resolution (down to $\lesssim$1\,\us) depending on the scientific case (e.g., millisecond pulsars). Of the 177 pulsars shown in Fig.~\ref{fig:mwapulsars} (and summarised in Figs.~\ref{fig:smartprofspage1} and \ref{fig:nonsmartprofspage1}, and Tables~\ref{tab:SMARTknownPSR} and ~\ref{tab:nonSMARTknownPSR}), the detection of 96 was facilitated by the SMART survey data, i.e. doubling of the number of pulsars detected with the MWA. For 34 pulsars, these are
also the first low-frequency detections ($\lesssim$300\,MHz); 30 of our pulsars have also been detected by the LOFAR \citep{bilous2016}. Collectively, these will help extend the low-frequency coverage of pulsars, providing a useful sample for further detailed studies of pulsar emission and/or population analysis. 

For many pulsars detected in SMART observations (Fig.~\ref{fig:smartprofspage1}), there have been multiple detections, by virtue of our sky tessellation strategy (Fig.~\ref{fig:smartprogress}), where the adjacent SMART pointings overlap $\sim$10\deg{} in R.A. and $\sim$12\deg{} in declination.
In most such cases, we have selected the best detection. The number of phase bins $N_{\rm bin}$ is chosen
to optimise  $\rm (S/N)_{\rm det}$ while retaining the ability to discern main
profile features, and varies from 64 to 1024.
For MSPs with short periods ($P \lesssim 5$\,ms),  $ N_{\rm bin} \approx P_{\rm ms}/0.1 $, where $P_{\rm ms}$ is the period in ms.

\begin{figure*}[tbh]
\begin{center}
\includegraphics[width=0.64\textwidth]{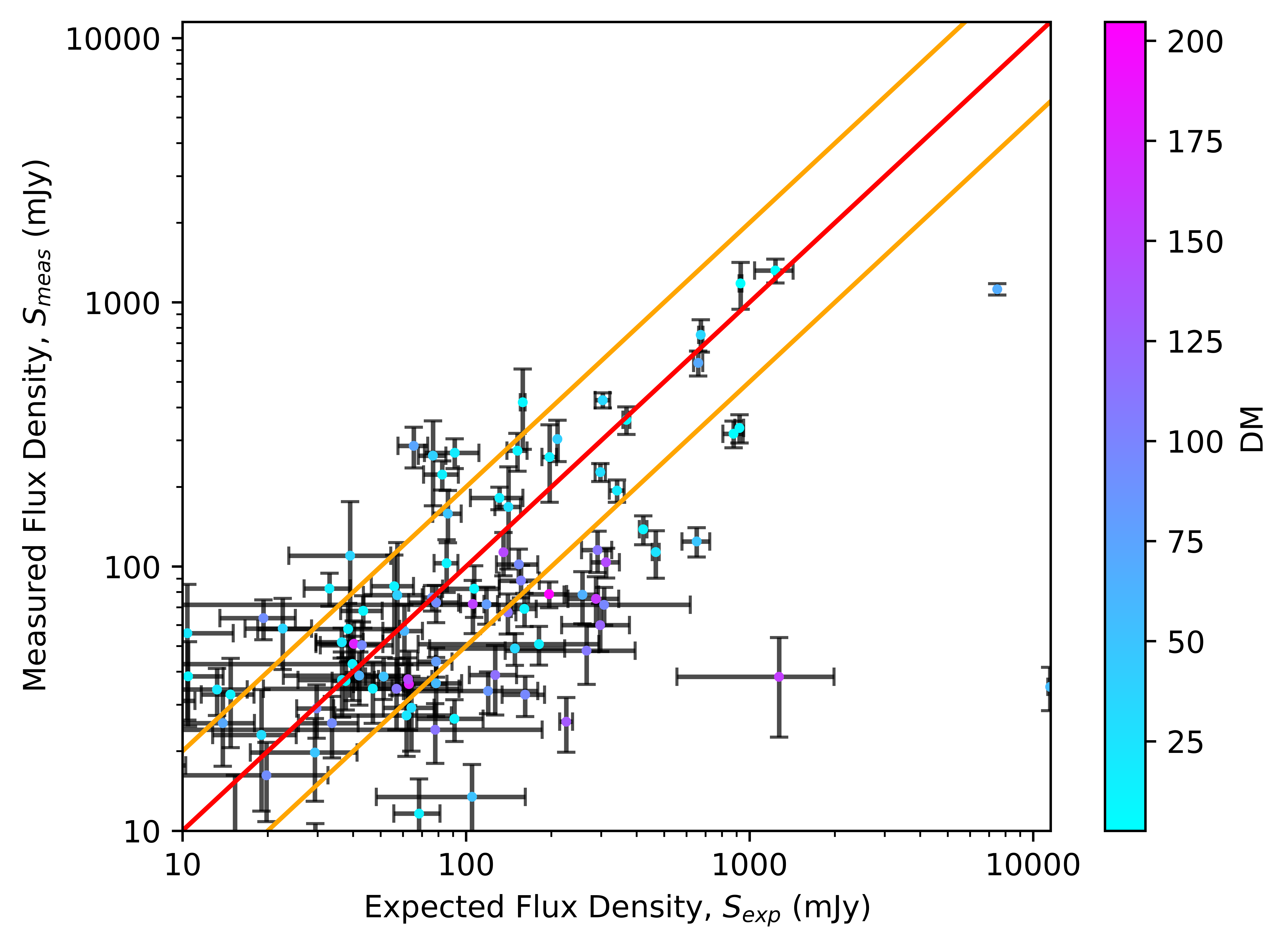}
\caption{
Measured flux densities $S_{\rm meas}$ against expected flux densities $S_{\rm exp}$ for 100 re-detections from the SMART survey processing; the measured values are corrected for the pulsar's offset relative to the centre of the primary beam, and can be as much as a factor of three. The expected values are extrapolations based on spectral fitting on the published flux density measurements of the pulsars (see text for details). The red line indicates where the two flux densities are equal whereas the grey-shaded region indicates a factor of two difference. 
}
\label{fig:fluxdensities}
\end{center}
\end{figure*}

\subsubsection{Detection sensitivity and flux densities}
\label{sec:fluxdensities} 

The re-detections from SMART are used to ascertain the survey sensitivity at 150\,MHz by comparing the measured pulsar flux densities \Smeas{} with their expected values \Sexp. To estimate \Smeas, folded profile data were flux density calibrated using the method outlined in \citet{meyers2017}. In brief, this makes use of the array factor formalism whereby the tied-array beam patterns are simulated by modelling them as the product of the tile beam pattern (primary beam response) and the array factor (in this case for the compact configuration of the Phase II array). This analysis yields estimates of the system temperature (\Tsys) and gain ($G$), which are then used to estimate the system equivalent flux density,  
${\rm SEFD} = \Tsys / G$. The simulations also naturally take into account the strong directional dependencies of these parameters. The variability in \Tsys{} and G over the duration of observation is accounted for as a source of error in the flux density estimate.

The expected flux densities \Sexp{} are essentially based on the extrapolation from the spectral forms deduced using the published flux density measurements of pulsars. This involved the use of a large body of flux density measurements made available by \citet{jankowski2018} as well as a number of other past publications. The model fits to the spectra are based on the algorithms described in the work of Jankowski et al. that employs the Akaike information criterion (AIC) to determine the best spectral model. Further details on the database repository and software description are presented in \citet{pulsarspectra}. 
The uncertainties in \Sexp{} thus largely reflect the quality and uncertainty of the model fit to the spectra (e.g., the error on the spectral index, or other spectral characteristics). An important caveat is that the bulk of the flux density measurements are from observations at frequencies $\gtrsim$400\,MHz and as a result may likely over-predict the expected values of \Sexp{} for pulsars that exhibit turnover, or a spectral break, or flattening at low frequencies.

A comparison of the measured flux densities of detected pulsars and the expected values as described above is shown in Fig.~\ref{fig:fluxdensities}. This includes 100 pulsars for which 
reasonable estimates of \Sexp{} were possible (i.e., four or more measurements available to allow meaningful estimates of the spectral form and the index $\alpha$), which limited the analysis to $\sim$80\% of the detected pulsars. 

As shown in the figure, within the caveats of the uncertainties associated with \Smeas{} and \Sexp{} (e.g., from beam simulations, 
and the spectral model fits) and the assumptions involved (e.g., spectral extrapolation to 150 MHz), we find there is reasonable agreement, with the measured  flux densities within a factor of two of their expected values. 
Evidently there are a small number of outliers where the analysis tends to over-predict \Sexp, and this may be to do with a spectral extrapolation error. In the case of some high DM pulsars, it is also possible there may be an underestimation of \Smeas, especially in
the cases where the scatter broadening is substantial. The ratio between the measured and expected values is \Smeas/\Sexp = $1.05 ^{+0.42} _{-0.31}$ (after excluding a 
small number of outlier points where  the ratio $>$10 or $<$0.1). 
Given this, and with the caveat that in general large flux density variabilities expected at low frequencies (as much as by a factor of $\sim$2-3 for moderate to high-DM pulsars, and even larger for low-DM ones), this provides a good demonstration that our SMART data processing is attaining a detection sensitivity in line with the expectations. 

The measured flux densities of detected pulsars also validate the minimum expected sensitivity (\Smin) for the shallow pass of our
survey, where $\Smin \sim 10$\,mJy for a 10$\sigma$ detection of long-period pulsars (see Fig.~\ref{fig:fluxdensities}). 
As seen from the figure, the faintest pulsar detections are $\Smeas \sim$7-8\,\mJy. 
In general, the measurements of low to moderate-DM pulsars are in better agreement compared to high-DM ones (i.e., $\gtrsim$50\,\dmu). 
Further details on the data used for the analysis, and the software description, are available in \citet{pulsarspectra}.

\subsubsection{Data release}
\label{sec:datarelease} 

Considering the inherent difficulties in accessing and processing VCS recorded data for pulsar detections and searching, we will make beamformed data of re-detected pulsars publicly accessible via a dedicated database repository (to be served by Data Central). We will adopt the preferred data format by other large pulsar databases \citep[e.g.,][]{hobbs2011}, i.e., folded archives in  10-s sub-integrations, 
with all 3072 channels and in full polarisation (i.e. four Stokes). For millisecond pulsars, we will provide coherently-dedispersed profiles in 1024 or more phase bins, in 24 coarse channels (1.28\,MHz). Our intent is a continual growth of this database as further detections accrue from the ongoing and future data processing. Given the  paucity of low-frequency pulsar detections in the southern sky, we hope this will become a useful resource, and potentially a reference catalogue of low-frequency pulsar detections in the SKA-Low era. 
Additionally, SMART survey raw data from 2018-2019 will be made available via a data retention project supported by Australian Research Data Commmons.\footnote{See \url{https://doi.org/10.25917/qd25-pg50} for data access.}

%% file: table-imaging-summary-J1357
\begin{table*}[!t]
\begin{threeparttable}
\caption{Localisation summary of PSR~\psrfour} \label{tab:psrfour}
\begin{tabular}{lccc}
\hline
% % % % Column names % % % %
  \multicolumn{1}{l}{Telescope/System} &
  \multicolumn{1}{c}{Method} &
  \multicolumn{1}{c}{Right Ascension (J2000)} &
  \multicolumn{1}{c}{Declination (J2000)} 
\\
\hline  
MWA VCS & Tied-array beams & $13^h57^m34^s$ & $-25\deg 33^\arcmin 11^\arcsec$  \\
uGMRT Band 3  & Imaging & $13^h57^m24.40^s$ & $-25\deg 30^\arcmin 39.0^\arcsec$ \\
uGMRT Band 4  & Imaging & $13^h57^m24.23^s$ & $-25\deg 30^\arcmin 38.7^\arcsec$ \\
Final position & Imaging & 13$^h$57$^m$24$^s$.3(2) & $-$25$\deg$30$^{\arcmin}$38$^{\arcsec}$.9(4) \\ 
\hline
\end{tabular}
\end{threeparttable}
\end{table*}

%%%%%%%%%%%%%%%%%%%%%%%%%%%%%%%%%%%%%%%%%%%%%%%%%%%%%%%%%%%%%%%%%%%%%%

%% file: 02summary-and-conclusions
\section{Summary and conclusions }
\label{sec:summary} 

The SMART survey, enabled by the advent of the Phase II upgrade of the MWA, is an ambitious all-sky project to search for pulsars and fast transients at low frequencies.  It exploits the  large FoV and voltage capture functionality of the MWA, the combination of which brings a survey efficiency of $\sim$450\,\sqdegphr, but at the expense of large data rates of 28\,\TBhr, and consequently, substantial computational costs in beamforming and search processing. The execution of the survey relies on the availability of the compact configuration of the Phase II array. Through a series of four dedicated campaigns over the past few years, the data collection part of the survey is now $\sim$75\% completed. 

In the first-pass processing as described in Paper I, where  10\,min of data from each observation (out of 80 min) are processed (in 2358 trial DMs, out to a maximum DM of 250\,\dmu), effectively a shallow survey is performed, reaching a sensitivity about one-third of  that will eventually be attainable with the planned deep-pass processing. Already three new pulsars have been found from the initial processing, where $\sim$3\% of the candidates have been scrutinised for promising ones. This initial processing has also led to re-detection of 120 previously known pulsars, thus bringing the total tally of MWA-detected pulsars to 180, i.e. more than a tripling of the number of pulsars detected with the MWA before the advent of the SMART survey. 

The voltage recording strategy employed for the survey enables a multitude of avenues for follow-ups and confirmations, including improved detection and localisation to sub-arcminute position, thereby facilitating prompt follow-ups using sensitive telescopes such as Parkes and uGMRT that operate at $\gtrsim$300 MHz. Not only that this 
 strategy is enabling an accelerated convergence to initial pulsar timing solution; e.g., by taking advantage of high-resolution ($\sim$arcsecond) imaging with the uGMRT, archival VCS observations (where available) can also be potentially exploited to obtain a full coherent timing solution, as demonstrated in this paper. 
 
 The early discoveries and science from the SMART provide an excellent demonstration of the potential benefits of using a next-generation survey instrument such as the MWA, in tandem with current-generation instruments (e.g., Parkes, uGMRT) for fruitful scientific exploitation. For example, PSR \psrone, the first MWA-discovered pulsar turned out to be a low-luminosity object  \citep{psrone}, and a promising target for probing the high-$b$ turbulent plasma. PSR~\psrtwo, the second pulsar discovered, exhibits intriguing sub-pulse drifting patterns and has a large nulling fraction of $\sim$77\% \citep{mcsweeney2022}. Follow-up studies of these pulsars are in progress with the uGMRT and FAST. The third pulsar, PSR~\psrthree, another potential low-luminosity object, is currently being followed up using the uGMRT and Parkes, as well as with the MWA. Finally, for PSR~\psrfour, a pulsar originally reported in the literature with an incorrect name and DM, we have obtained a precise position via uGMRT follow-up to facilitate a high-resolution imaging and source identification. 

The initial discoveries and results presented in this paper hint at a promising future for the SMART. As efforts around data processing and candidate scrutiny ramp up over the coming years,
we anticipate the discovery rate to increase. 
The SMART is well positioned to emerge as a low-frequency version of the HTRU survey with Parkes, and as an important reference for even more ambitious surveys planned with the SKA-Low.

%% file: 02acknowledgements
\begin{acknowledgement}

We thank an anonymous referee for several useful comments that helped to improve the content and presentation of this paper. 
%MRO/Pawsey
The scientific work made use of 
%\textbf{Inyarrimanha Ilgari Bundara, the CSIRO Murchison Radio-astronomy Observatory}. 
Inyarrimanha Ilgari Bundara, the CSIRO Murchison Radio-astronomy Observatory. 
We acknowledge the Wajarri Yamaji people as the traditional owners of the Observatory site. This work was supported by resources provided by the Pawsey Supercomputing Centre with funding from the Australian Government and the Government of Western Australia.
%OzSTAR
This work was supported by resources awarded under Astronomy Australia Ltd's ASTAC merit allocation scheme on the OzSTAR national facility at the Swinburne University of Technology. The OzSTAR program receives funding in part from the Astronomy National Collaborative Research Infrastructure Strategy (NCRIS) allocation provided by the Australian Government.
%ADACS
%The development of SMART webapp was facilitated by the software support scheme of ADACS.
%, and we thank Rebecca Lange and Simon O'Toole for their contributions to this project.
%Facilities
%GMRT
The GMRT is run by the National Centre for Radio Astrophysics
of the Tata Institute of Fundamental Research, India. 
%Parkes
The Parkes radio telescope is part of the Australia Telescope National Facility which is funded by the Australian Government for operation as a National Facility managed by CSIRO.
%Other 
%We thank L. Levin for help with the simulation analysis of MSPs. 

%Software
\textit{Software}: We acknowledge the use of the following software/packages for this work: CASA \citep{casa}, DSPSR \citep{dspsr_ascl, dspsr}, PRESTO \citep{presto, presto_ascl}, PSRCHIVE \citep{psrchive,psrchive_ascl}, Tempo2 \citep{tempo2_1, tempo2_ascl}, PINT \citep{pint_ascl, pint}, PsrPopPy \citep{Bates2014}, Nextflow \citep{nextflow}.
%}

\end{acknowledgement}

%% file: figures
\begin{figure*}[p]
\begin{center}
\includegraphics[height=237mm,angle=0]{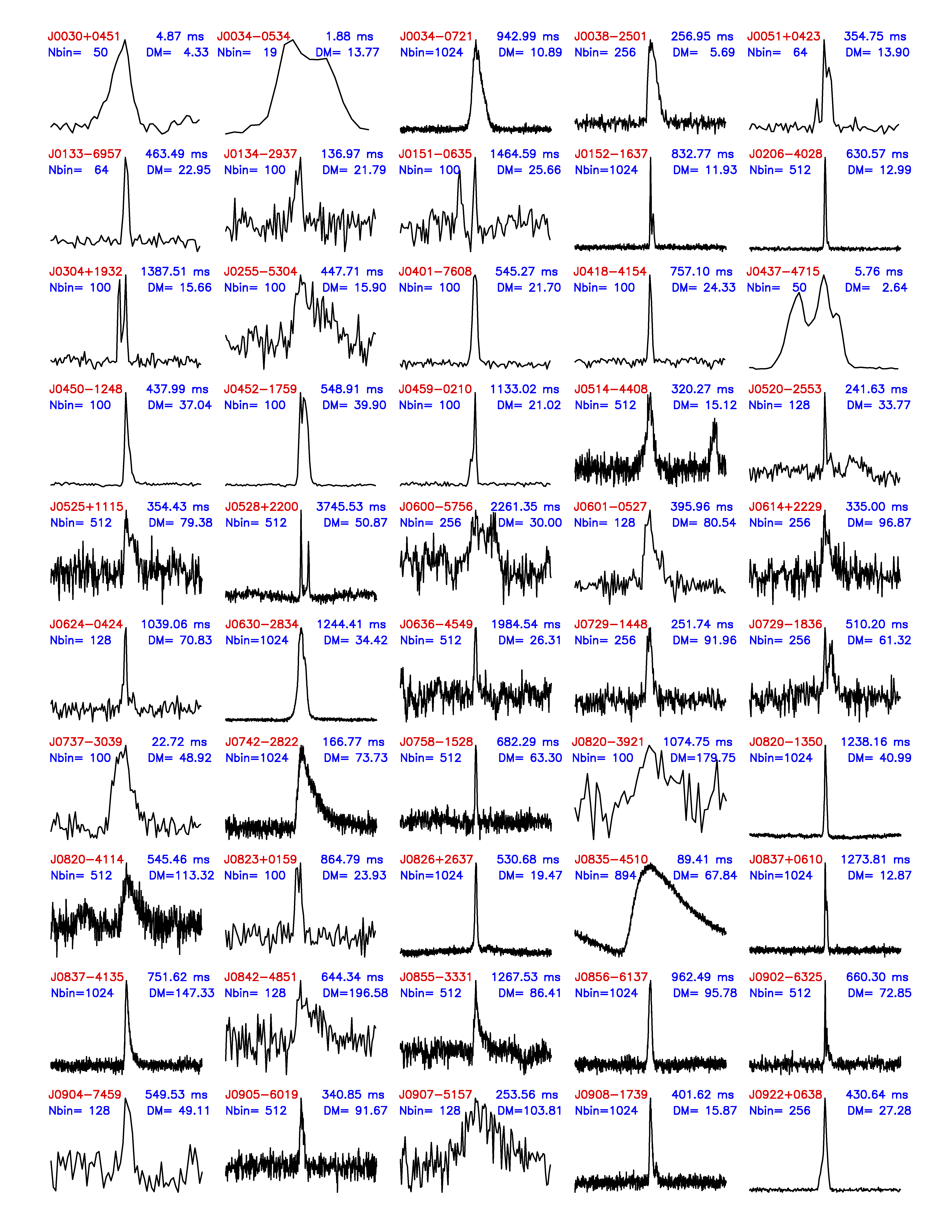}
\caption{
Integrated pulse profiles for 123 re-detected pulsars in the SMART data processing. The period, DM, and the number of phase bins are shown in each panel. All detections were made in the SMART survey band 140-170\,MHz. 
}
\label{fig:smartprofspage1}
\end{center}
\end{figure*}

\begin{figure*}[p]
\begin{center}
 \continuedfloat
\includegraphics[height=237mm,angle=0]{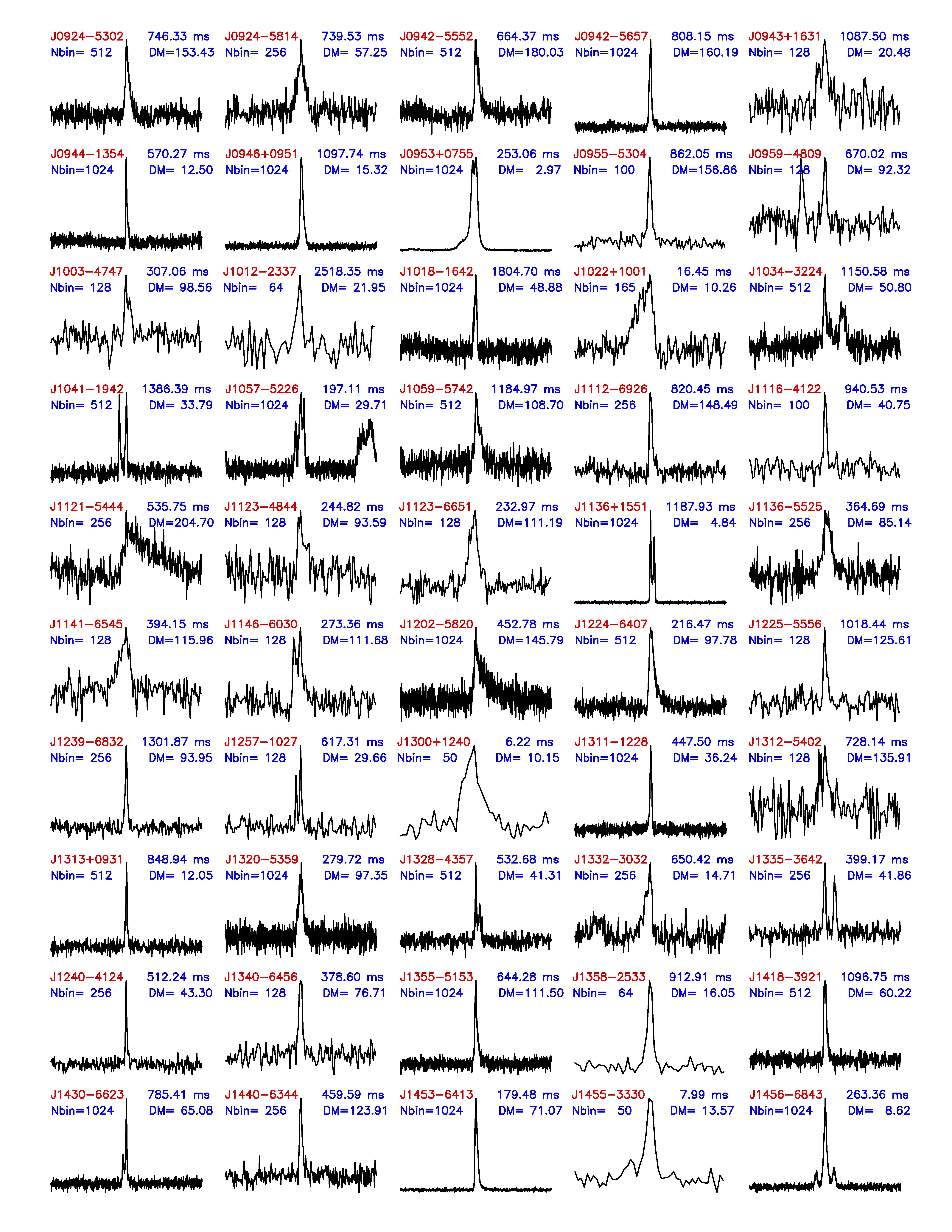}
\caption{continued.}
%\caption{
%Integrated pulse profiles for 123 re-detected pulsars in the SMART data processing. The period, DM, and the number of phase bins are shown in each panel. All detections were made in the SMART survey band 140-170 MHz. }
\label{fig:smartprofspage2}
\end{center}
\end{figure*}

\begin{figure*}[p]
\begin{center}
 \continuedfloat
\includegraphics[height=237mm,angle=0]{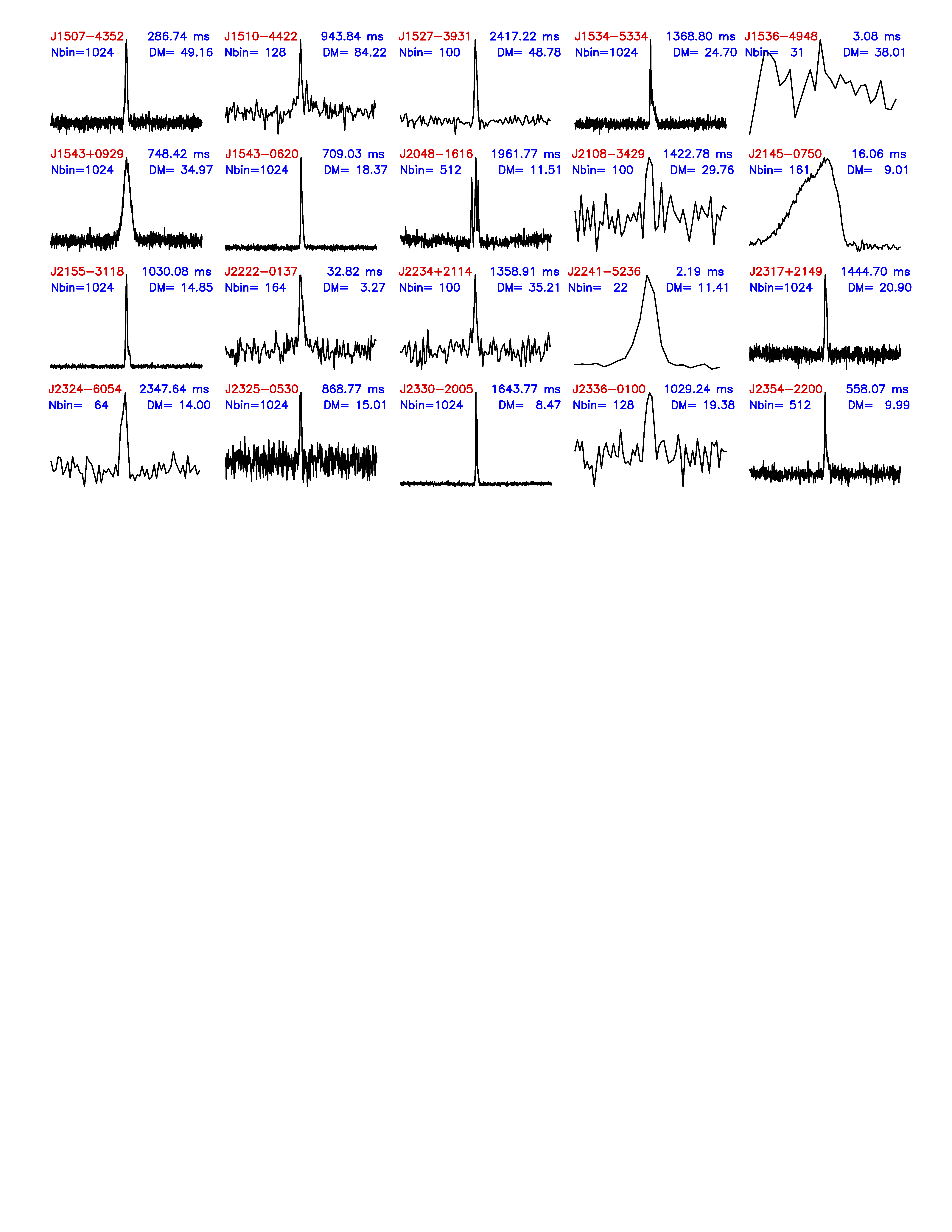}
\caption{continued.}
%\caption{
%Integrated pulse profiles for 123 re-detected pulsars in the SMART data processing. The period, DM, and the number of phase bins are shown in each panel. All detections were made in the SMART survey band 140-170 MHz. }
\label{fig:smartprofspage3}
\end{center}
\end{figure*}

%%%%%%%%TABLE of SMART pulsar detections 
% Long table Mengyao is putting together
% SMART pulsars - version as of 28May22
%\input{smart-table-28may22}
%fixed up for the decimal points RB
\input{smart-table-27sept22}

%nonSMART pulsars version as of 28May22
\input{non-smart-table-28may22}

\begin{figure*}[p]
\begin{center}
\includegraphics[height=237mm,angle=0]{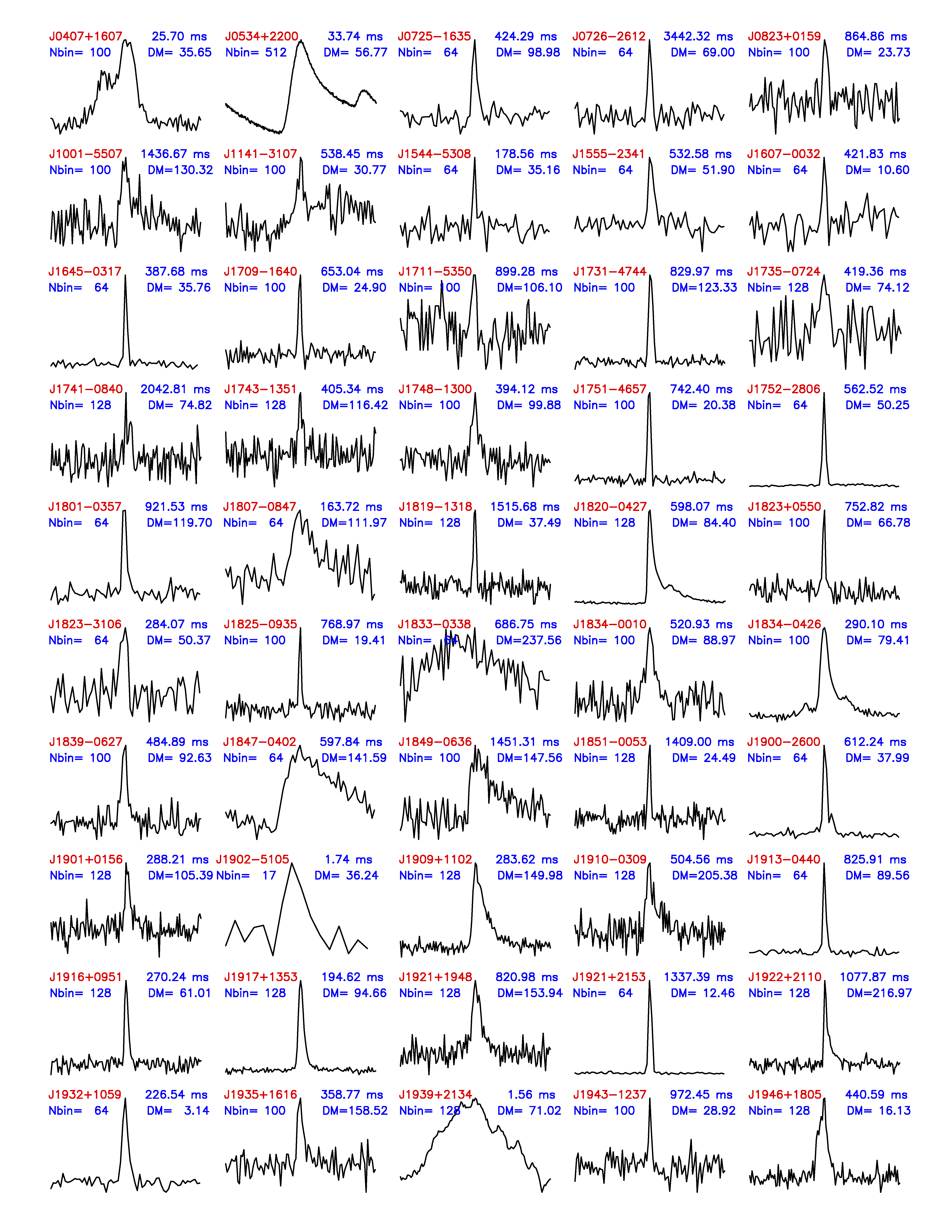}
\caption{
Integrated pulse profiles for 58 pulsars detected in observations of various non-SMART projects. The period, DM, and the number of phase bins are shown in each panel. Th majority of these detections were made in the 170-200\,MHz band, while a few in the SMART survey band (140-170\,MHz). 
}
\label{fig:nonsmartprofspage1}
\end{center}
\end{figure*}

\begin{figure*}[p]
\begin{center}
    \continuedfloat
\includegraphics[height=237mm,angle=0]{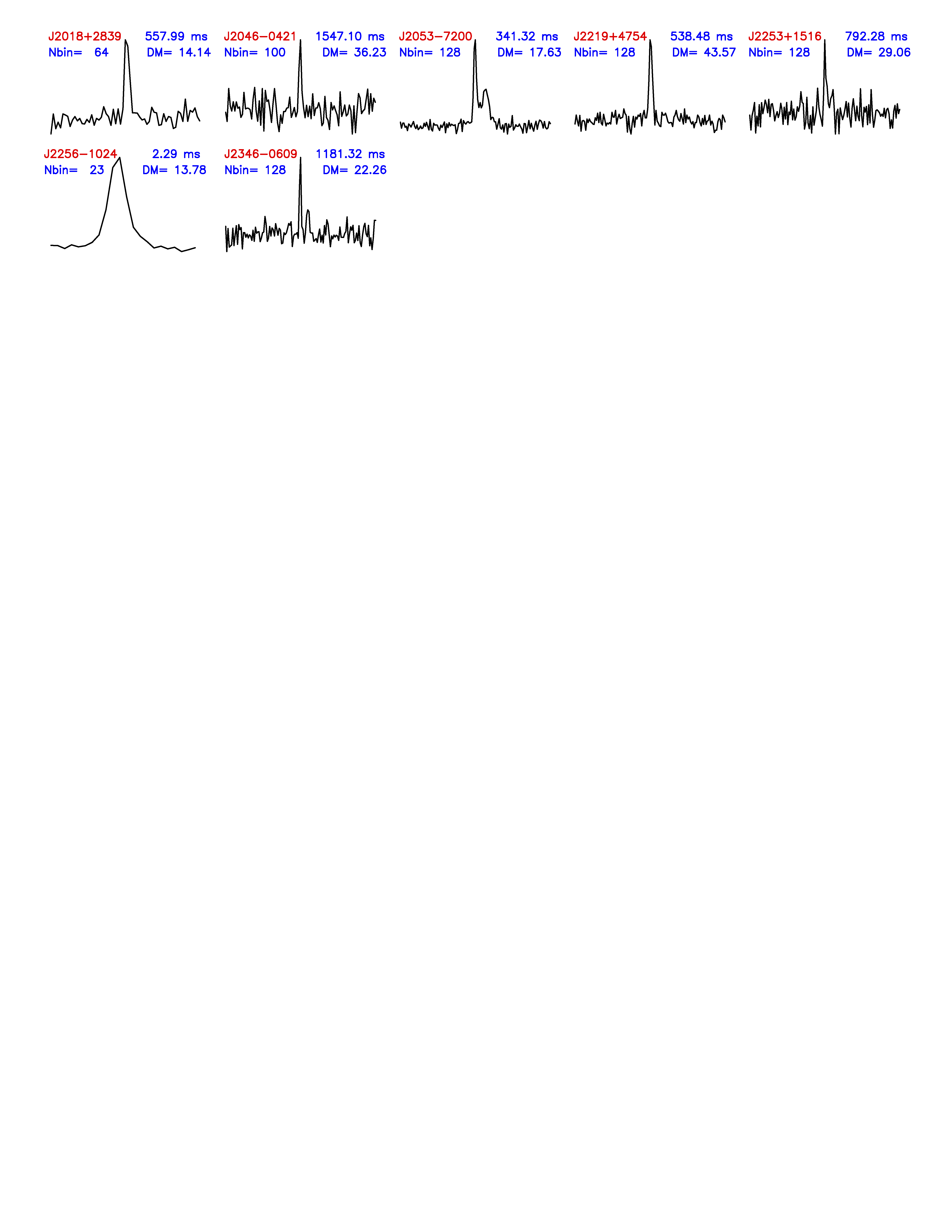}
   \caption{continued.}
%\caption{
%Integrated pulse profiles for 58 pulsars detected in observations of various non-SMART projects. The period, DM, and the number of phase bins are shown in each panel. Th majority of these detections were made in the 170-200 MHz band, while a few in the SMART survey band (140-170 MHz). 
%}
\label{fig:nonsmartprofspage2}
\end{center}
\end{figure*}

%% file: smart-table-27sept22
\onecolumn

\begin{center}
\begin{ThreePartTable}
\begin{TableNotes}  %%% This should come before longtable and it is not tablenotes but TableNotes
\footnotesize
    % \item[1] Signal-to-noise ratio for MWA tied-array detection.
    % \item[2] Flux density for MWA tied-array detection at 155\,MHz.
    \item[1] Target offset from observation centre. 
    \item[2] Low-frequency ($<400$\,MHz) information available in the literature: `M' for MWA-incoh/MWA-image \citep{bell2016,Murphy2017,xue2017,kaur2019}, `L' for LOFAR \citep{Kondratiev2016,Bilous2016,sanidas2019,Bondonneau2020}, `W' for LWA \citep{Stovall2015}, `G' for GMRT \citep{Bhattacharyya2016,Frail2016}, `g' for GBT-GBNCC \citep{McEwen2020}, `E’ for Effelsberg \citep{Sieber1973}, `P' for Pushchino \citep{Izvekova1981,Malofeev2000,Kuzmin2001,Tyul'bashev2016},  `U' for Ukrainian T-shaped Radio telescope-UTR2 \citep{Zakharenko2013}.
    % BWM: we should really use the actual citations here I think...
\end{TableNotes}
  
\begin{longtable}{lrrrrcccc}
\caption{Known pulsars detected in SMART survey observations.} \label{tab:SMARTknownPSR} \\
\hline
% % % % Column names % % % %
  \multicolumn{1}{c}{Pulsar} &
  \multicolumn{1}{c}{Period} &
  \multicolumn{1}{c}{DM} &
  \multicolumn{1}{c}{Offset\tnote{1}} &
%   \multicolumn{1}{c}{S/N\tnote{1}} &
%   \multicolumn{1}{c}{Flux\tnote{2}} &
  \multicolumn{1}{c}{S/N} &
  \multicolumn{1}{c}{Mean flux density} &
  \multicolumn{1}{c}{Obs ID} &
  \multicolumn{1}{c}{Notes\tnote{2}} \\
% % % % Column units % % % %
  \multicolumn{1}{c}{} &
  \multicolumn{1}{c}{(\ms)} &
  \multicolumn{1}{c}{(\dmu)} &
  \multicolumn{1}{c}{($\rm deg$)} &
  \multicolumn{1}{c}{} &
  \multicolumn{1}{c}{(\mJy)} &
  \multicolumn{1}{c}{} &
  \multicolumn{1}{c}{} \\
\hline
\endfirsthead

\hline
% % % % Column names % % % %
  \multicolumn{1}{c}{Pulsar} &
  \multicolumn{1}{c}{Period} &
  \multicolumn{1}{c}{DM} &
  \multicolumn{1}{c}{Offset\tnote{1}} &
%   \multicolumn{1}{c}{S/N\tnote{1}} &
%   \multicolumn{1}{c}{Flux\tnote{2}} &
  \multicolumn{1}{c}{S/N} &
  \multicolumn{1}{c}{Mean flux density} &
  \multicolumn{1}{c}{Obs ID} &
  \multicolumn{1}{c}{Notes\tnote{2}} \\
% % % % Column units % % % %
  \multicolumn{1}{c}{} &
  \multicolumn{1}{c}{(\ms)} &
  \multicolumn{1}{c}{(\dmu)} &
  \multicolumn{1}{c}{($\rm deg$)} &
  \multicolumn{1}{c}{} &
  \multicolumn{1}{c}{(\mJy)} &
  \multicolumn{1}{c}{} &
  \multicolumn{1}{c}{} \\
\hline
\endhead

\hline
\multicolumn{9}{r}{{Continued on next page}} \\
\endfoot

\hline
\insertTableNotes
\endlastfoot

J0030+0451 & 4.865 & 4.33 & 4.3 & 6.3 & 37$\pm$8 & 1255444104 &  G, L, g, P \\
J0034-0534 & 1.877 & 13.77 & 7.3 & 3.6 & 260$\pm$85 & 1255444104 &  M, G, L, g, P \\
J0034-0721 & 942.951 & 10.92 & 9.0 & 64.0 & 692$\pm$42 & 1255444104 &  P, E, W, U, M, L \\
J0038-2501 & 256.926 & 5.71 & 8.5 & 24.6 & 44$\pm$6 & 1226062160 &  g \\
J0051+0423 & 354.732 & 13.93 & 14.6 & 26.4 & 59$\pm$17 & 1225118240 &  U, L, g \\
J0133-6957 & 463.474 & 22.95 & 7.7 & 25.1 & 12$\pm$4 & 1227009976 &   \\
J0134-2937 & 136.962 & 21.81 & 20.7 & 10.6 & 11$\pm$3 & 1226062160 &   \\
J0151-0635 & 1464.665 & 25.66 & 17.3 & 13.9 & 8$\pm$3 & 1252177744 &  P, U, G \\
J0152-1637 & 832.742 & 11.93 & 11.5 & 74.4 & 131$\pm$17 & 1225462936 &  W, P, G, L \\
J0206-4028 & 630.551 & 12.9 & 11.1 & 68.2 & 1289$\pm$16 & 1224859816 &  M \\
J0255-5304 & 447.708 & 15.9 & 4.4 & 9.7 & 67$\pm$10 & 1253471952 &   \\
J0304+1932 & 1387.584 & 15.66 & 12.7 & 22.3 & 35$\pm$9 & 1254594264 &  P, W, L \\
J0401-7608 & 545.254 & 21.7 & 4.9 & 52.0 & 74$\pm$18 & 1255803168 &   \\
J0418-4154 & 757.119 & 24.33 & 3.9 & 42.3 & 38$\pm$11 & 1253991112 &  M, G \\
J0437-4715 & 5.757 & 2.64 & 15.4 & 6.2 & 1179$\pm$238 & 1257617424 &  M \\
J0450-1248 & 438.014 & 37.04 & 14.8 & 81.3 & 131$\pm$33 & 1256407632 &  P \\
J0452-1759 & 548.939 & 39.9 & 9.1 & 37.0 & 186$\pm$14 & 1257010784 &  P, M \\
J0459-0210 & 1133.076 & 21.02 & 3.9 & 81.5 & 68$\pm$22 & 1256407632 &  L, g \\
J0514-4408 & 320.271 & 15.12 & 13.3 & 13.2 & 91$\pm$8 & 1257617424 &  G \\
J0520-2553 & 241.642 & 33.77 & 3.8 & 20.5 & 18$\pm$5 & 1257010784 &  g \\
J0525+1115 & 354.438 & 79.42 & 12.8 & 8.7 & 44$\pm$6 & 1259685792 &  P, L \\
J0528+2200 & 3745.539 & 50.87 & 10.0 & 31.6 & 38$\pm$7 & 1259685792 &  L, P, E, W, P \\
J0600-5756 & 2261.365 & 30.0 & 2.9 & 8.6 & 27$\pm$3 & 1257617424 &   \\
J0601-0527 & 395.969 & 80.54 & 7.7 & 19.7 & 72$\pm$12 & 1259427304 &  P \\
J0614+2229 & 334.96 & 96.91 & 4.0 & 9.7 & 33$\pm$6 & 1259685792 &  P, L \\
J0624-0424 & 1039.076 & 70.84 & 16.3 & 13.4 & 26$\pm$8 & 1260638120 &  P, L \\
J0630-2834 & 1244.419 & 34.43 & 12.2 & 126.3 & 1353$\pm$91 & 1261241272 &  E, W, P, M, L \\
J0636-4549 & 1984.597 & 26.31 & 7.3 & 7.1 & 19$\pm$6 & 1258221008 &   \\
J0729-1448 & 251.659 & 91.89 & 12.6 & 12.9 & 64$\pm$11 & 1266155952 &  g \\
J0729-1836 & 510.16 & 61.29 & 13.6 & 12.4 & 72$\pm$11 & 1266155952 &   \\
J0737-3039A & 22.699 & 48.92 & 4.9 & 22.2 & 173$\pm$28 & 1261241272 &  M, L, g \\
J0742-2822 & 166.762 & 73.73 & 23.0 & 11.1 & 362$\pm$40 & 1265470568 &  P, M \\
J0758-1528 & 682.265 & 63.33 & 6.0 & 21.2 & 39$\pm$9 & 1266155952 &   \\
J0820-1350 & 1238.13 & 40.94 & 0.9 & 112.7 & 411$\pm$48 & 1266155952 &  P \\
J0820-3921 & 1073.567 & 179.4 & 1.0 & 5.8 & 51$\pm$11 & 1265983624 &  W, P, M, L \\
J0820-4114 & 545.446 & 113.4 & 20.0 & 7.3 & 34$\pm$7 & 1265470568 &  g \\
J0823+0159 & 864.873 & 23.73 & 14.9 & 11.1 & 37$\pm$10 & 1266155952 &  M \\
J0826+2637 & 530.661 & 19.48 & 8.1 & 51.0 & 194$\pm$19 & 1265725128 &  L, P, E, W, U, M \\
J0835-4510 & 89.328 & 67.77 & 21.8 & 23.0 & 1121$\pm$55 & 1266680784 &  E, M \\
J0837+0610 & 1273.768 & 12.86 & 13.2 & 61.2 & 138$\pm$17 & 1265725128 &  L, P, E, W, P, M, U \\
J0837-4135 & 751.625 & 147.2 & 13.4 & 24.4 & 144$\pm$12 & 1266329600 &  M \\
J0842-4851 & 644.354 & 196.85 & 6.1 & 6.4 & 43$\pm$8 & 1266329600 &   \\
J0855-3331 & 1267.536 & 86.64 & 9.6 & 15.7 & 76$\pm$9 & 1265470568 &  M \\
J0856-6137 & 962.511 & 95.0 & 10.4 & 32.9 & 93$\pm$9 & 1266932744 &  M \\
J0902-6325 & 660.313 & 72.72 & 8.7 & 22.7 & 23$\pm$5 & 1266932744 &   \\
J0904-7459 & 549.554 & 51.1 & 3.4 & 6.9 & 13$\pm$4 & 1266932744 &   \\
J0905-6019 & 340.854 & 91.4 & 11.8 & 14.8 & 32$\pm$5 & 1266932744 &   \\
J0907-5157 & 253.558 & 103.72 & 4.7 & 7.1 & 89$\pm$13 & 1266329600 &  M, G \\
J0908-1739 & 401.626 & 15.88 & 20.6 & 20.2 & 62$\pm$6 & 1267283936 &  P, W, L \\
J0922+0638 & 430.627 & 27.3 & 5.0 & 45.0 & 184$\pm$17 & 1264867416 &  L, P, M, U \\
J0924-5302 & 746.338 & 152.9 & 17.1 & 14.1 & 64$\pm$9 & 1266680784 &  M, G \\
J0924-5814 & 739.505 & 57.4 & 14.4 & 10.2 & 30$\pm$5 & 1266932744 &   \\
J0942-5552 & 664.389 & 180.16 & 17.2 & 14.4 & 37$\pm$5 & 1266932744 &  M \\
J0942-5657 & 808.164 & 159.74 & 16.2 & 44.4 & 95$\pm$11 & 1266932744 &  M \\
J0943+1631 & 1087.418 & 20.34 & 11.7 & 11.1 & 34$\pm$7 & 1267111608 &  P, U, L \\
J0944-1354 & 570.264 & 12.5 & 11.6 & 24.8 & 27$\pm$5 & 1267283936 &  P \\
J0946+0951 & 1097.706 & 15.32 & 14.2 & 53.4 & 182$\pm$18 & 1267111608 &  L, P, E, U \\
J0953+0755 & 253.065 & 2.97 & 14.3 & 96.9 & 775$\pm$51 & 1267111608 &  L, P, E, U, M \\
J0955-5304 & 862.122 & 156.9 & 14.2 & 35.3 & 49$\pm$13 & 1266680784 &   \\
J0959-4809 & 670.086 & 92.7 & 9.8 & 19.0 & 110$\pm$8 & 1266680784 &  M \\
J1003-4747 & 307.074 & 98.49 & 9.1 & 17.6 & 29$\pm$7 & 1266680784 &  G \\
J1012-2337 & 2517.945 & 22.51 & 19.4 & 14.3 & 39$\pm$5 & 1268321832 &  M \\
J1018-1642 & 1804.695 & 48.82 & 20.8 & 13.1 & 16$\pm$5 & 1268321832 &  P \\
J1022+1001 & 16.453 & 10.25 & 16.4 & 10.7 & 68$\pm$10 & 1264867416 &  M, L, P \\
J1034-3224 & 1150.59 & 50.75 & 14.9 & 12.9 & 73$\pm$7 & 1268321832 &  g \\
J1041-1942 & 1386.368 & 33.78 & 14.6 & 24.5 & 49$\pm$7 & 1268321832 &   \\
J1057-5226 & 197.115 & 29.69 & 4.5 & 19.1 & 287$\pm$15 & 1267459328 &  M \\
J1059-5742 & 1185.003 & 108.7 & 17.5 & 21.6 & 49$\pm$10 & 1301240224 &   \\
J1112-6926 & 820.488 & 148.4 & 7.5 & 21.5 & 38$\pm$8 & 1301240224 &  M \\
J1116-4122 & 943.158 & 40.53 & 13.6 & 16.7 & 54$\pm$8 & 1267459328 &  M \\
J1121-5444 & 535.787 & 204.7 & 0.3 & 10.1 & 79$\pm$9 & 1267459328 &  M \\
J1123-4844 & 244.838 & 92.92 & 6.2 & 6.2 & 16$\pm$5 & 1267459328 &   \\
J1123-6651 & 232.976 & 111.2 & 8.3 & 11.3 & 40$\pm$8 & 1301240224 &   \\
J1136+1551 & 1187.913 & 4.84 & 14.1 & 151.6 & 318$\pm$37 & 1268063336 &  L, P, E, W, U, M \\
J1136-5525 & 364.713 & 85.11 & 1.9 & 7.4 & 34$\pm$6 & 1267459328 &   \\
J1141-6545 & 393.899 & 116.08 & 8.1 & 14.1 & 55$\pm$10 & 1301240224 &  M \\
J1146-6030 & 273.375 & 111.68 & 6.4 & 9.4 & 24$\pm$6 & 1267459328 &   \\
J1202-5820 & 452.803 & 145.41 & 14.1 & 13.9 & 88$\pm$11 & 1301240224 &   \\
J1224-6407 & 216.48 & 97.69 & 7.9 & 18.1 & 89$\pm$9 & 1301240224 &   \\
J1225-5556 & 1018.453 & 125.84 & 8.9 & 12 & 15$\pm$5.6 & 1267459328 &   \\
J1239-6832 & 1301.923 & 94.3 & 3.4 & 20.8 & 26$\pm$7 & 1301240224 &   \\
J1240-4124 & 512.242 & 44.1 & 1.1 & 22.2 & 34$\pm$9 & 1301412552 &  G \\
J1257-1027 & 617.308 & 29.63 & 5.0 & 18.0 & 29$\pm$9 & 1300809400 &  P \\
J1300+1240 & 6.219 & 10.17 & 15.6 & 9.4 & 34$\pm$9 & 1301847296 &  P, L, g \\
J1311-1228 & 447.518 & 36.21 & 16.4 & 26.1 & 51$\pm$9 & 1301847296 &  P \\
J1312-5402 & 728.154 & 133.0 & 15.8 & 8.5 & 26$\pm$6 & 1267459328 &   \\
J1313+0931 & 848.933 & 12.04 & 11.1 & 23.3 & 37$\pm$8 & 1301847296 &  L \\
J1320-5359 & 279.738 & 97.1 & 15.4 & 10.7 & 36$\pm$9 & 1301412552 &  G \\
J1328-4357 & 532.699 & 42.0 & 9.7 & 20.6 & 55$\pm$9 & 1301412552 &   \\
J1332-3032 & 650.434 & 15.1 & 4.7 & 12.1 & 51$\pm$9 & 1301674968 &  g \\
J1335-3642 & 399.192 & 41.82 & 11.6 & 13.2 & 49$\pm$9 & 1301412552 &  g \\
J1340-6456 & 378.622 & 76.99 & 8.9 & 12.1 & 24$\pm$7 & 1301240224 &   \\
J1355-5153 & 644.305 & 112.1 & 3.0 & 28.7 & 146$\pm$16 & 1302106648 &   \\
J1358-2533 & 912.971 & 16.05 & 3.3 & 26.0 & 27$\pm$8 & 1301674968 &  g \\
J1418-3921 & 1096.806 & 60.49 & 15.9 & 17.3 & 53$\pm$12 & 1302106648 &  g \\
J1430-6623 & 785.443 & 65.1 & 12.1 & 34.7 & 112$\pm$15 & 1302106648 &  M \\
J1440-6344 & 459.607 & 124.2 & 10.3 & 17.7 & 60$\pm$12 & 1302106648 &  M \\
J1453-6413 & 179.487 & 71.25 & 11.5 & 110.0 & 590$\pm$64 & 1302106648 &  M \\
J1455-3330 & 7.987 & 13.57 & 6.9 & 7.8 & 58$\pm$14 & 1302282040 &   \\
J1456-6843 & 263.377 & 8.61 & 15.3 & 51.1 & 433$\pm$36 & 1302106648 &  M \\
J1507-4352 & 286.758 & 48.7 & 4.8 & 25.1 & 125$\pm$16 & 1302282040 &  M \\
J1510-4422 & 943.871 & 84.0 & 5.5 & 13.2 & 51$\pm$13 & 1302282040 &   \\
J1527-3931 & 2417.605 & 48.8 & 7.3 & 24.0 & 31$\pm$13 & 1302282040 &  g \\
J1534-5334 & 1368.882 & 24.82 & 15.2 & 30.5 & 95$\pm$17 & 1302282040 &  M \\
J1536-4948 & 3.08 & 38.0 & 12.5 & 8.7 & 13$\pm$12 & 1302282040 &  \\
J1543+0929 & 748.448 & 34.98 & 15.8 & 26.5 & 426$\pm$28 & 1302540536 &  L, P, E, W, M \\
J1543-0620 & 709.064 & 18.3 & 14.6 & 69.0 & 270$\pm$35 & 1302712864 &  P, W, U, M, L, g \\
J2048-1616 & 1961.572 & 11.46 & 17.6 & 32.1 & 69$\pm$10 & 1222435400 &  P, M \\
J2108-3429 & 1423.102 & 30.22 & 11.7 & 7.5 & 13$\pm$5 & 1222435400 &  g \\
J2145-0750 & 16.052 & 9.0 & 17.7 & 9.2 & 61$\pm$11 & 1222697776 &  P, M, L \\
J2155-3118 & 1030.002 & 14.85 & 4.7 & 109.9 & 223$\pm$28 & 1222435400 &  M \\
J2222-0137 & 32.818 & 3.27 & 8.6 & 13.3 & 31$\pm$7 & 1221832280 &  L, g \\
J2234+2114 & 1358.745 & 35.08 & 5.5 & 5.2 & 9$\pm$7 & 1223042480 &  P, L, g \\
J2241-5236 & 2.187 & 11.41 & 12.0 & 27.9 & 705$\pm$37 & 1224252736 &  M \\
J2317+2149 & 1444.653 & 20.87 & 6.2 & 24.0 & 52$\pm$7 & 1223042480 &  L, P, U \\
J2324-6054 & 2347.488 & 14.0 & 11.8 & 17.0 & 12$\pm$4 & 1227009976 &   \\
J2325-0530 & 868.735 & 14.97 & 10.7 & 10.7 & 15$\pm$4 & 1222697776 &  g \\
J2330-2005 & 1643.622 & 8.46 & 9.7 & 112.4 & 87$\pm$14 & 1226062160 &  W, P, L \\
J2336-01 & 1029.8 & 19.6 & 15.3 & 11.7 & 53$\pm$10 & 1222697776 &   \\
J2354-22 & 557.996 & 9.9 & 4.2 & 25.2 & 16$\pm$3 & 1226062160 &  g \\
    \hline
\end{longtable}

\end{ThreePartTable}
\end{center}

%% file: non-smart-table-28may22
\onecolumn
\begin{center}

\begin{ThreePartTable}
\begin{TableNotes}  %%% This should come before longtable and it is not tablenotes but TableNotes
\footnotesize
    \item[1] Target offset from observation centre.
    \item[2] The mode of beamforming used. I: incoherent, where the detected powers of each tile are summed. C: coherent/tied-array, where the voltages of each tile are summed (after appropriate phase rotations to the pointing-direction) and then detected.
    For incoherent detections, flux density estimates are using the method outlined in \citet{xue2017}, whereas for coherent beam detections, they are estimated using the method described in \S~\ref{sec:redetections}. 
\end{TableNotes}

\begin{longtable}{crrrcrcc}
\caption{Known pulsars detected with MWA observations not associated with SMART.} \label{tab:nonSMARTknownPSR} \\
\hline
% % % % Column names % % % %
  \multicolumn{1}{c}{Pulsar} &
  \multicolumn{1}{c}{Period} &
  \multicolumn{1}{c}{DM} &
  \multicolumn{1}{c}{Offset\tnote{1}} &
  \multicolumn{1}{c}{Obs. freq.} &
  \multicolumn{1}{c}{S/N} &
  \multicolumn{1}{c}{Beamform\tnote{2}} &
  \multicolumn{1}{c}{Mean flux density} \\
% % % % Column units % % % %
  \multicolumn{1}{c}{} &
  \multicolumn{1}{c}{(\ms)} &
  \multicolumn{1}{c}{(\dmu)} &
  \multicolumn{1}{c}{($\rm deg$)} &
  \multicolumn{1}{c}{($\rm MHz$)} &
  \multicolumn{1}{c}{} &
  \multicolumn{1}{c}{type} &
  \multicolumn{1}{c}{(\mJy)} \\
\hline
\endfirsthead

\hline
% % % % Column names % % % %
  \multicolumn{1}{c}{Pulsar} &
  \multicolumn{1}{c}{Period} &
  \multicolumn{1}{c}{DM} &
  \multicolumn{1}{c}{Offset\tnote{1}} &
  \multicolumn{1}{c}{Obs. freq.} &
  \multicolumn{1}{c}{S/N} &
  \multicolumn{1}{c}{Beamform\tnote{2}} &
  \multicolumn{1}{c}{Mean flux density} \\
% % % % Column units % % % %
  \multicolumn{1}{c}{} &
  \multicolumn{1}{c}{(\ms)} &
  \multicolumn{1}{c}{(\dmu)} &
  \multicolumn{1}{c}{($\rm deg$)} &
  \multicolumn{1}{c}{($\rm MHz$)} &
  \multicolumn{1}{c}{} &
  \multicolumn{1}{c}{type} &
  \multicolumn{1}{c}{(\mJy)} \\
\hline
\endhead

\hline
\multicolumn{8}{r}{{Continued on next page}} \\
\endfoot

\hline
\insertTableNotes
\endlastfoot

    J0407+1607 & 25.702  & 35.65  & 3.4   & 155   & 20.1     & C      & 60$\pm$9  \\
    J0534+2200 & 33.392  & 56.77  & 0.1   & 155   & 538      & C      & 6450$\pm$1200 \\
    J0725-1635 & 424.311  & 98.98  & 17.4  & 155   & 16.3    & C      & 10$\pm$4  \\
    J0726-2612 & 3442.308  & 69.40  & 13.1  & 155   & 26.9   & C      & 10$\pm$4  \\
    J0823+0159 & 864.873  & 23.73  & 8     & 185   & 9.6     & I      & 35$\pm$5  \\
    J1001-5507 & 1436.631  & 130.32  & 1.4   & 185   & 6.8   & I      & 53$\pm$5  \\
    J1141-3107 & 538.432  & 30.77  & 2.7   & 185   & 10.8    & I      & 53$\pm$6  \\
    J1544-5308 & 178.554  & 35.16  & 7.6   & 185   & 8.8     & I      & 102$\pm$15  \\
    J1555-2341 & 532.578  & 51.90  & 21.8  & 185   & 29.5    & C      & 70$\pm$17  \\
    J1607-0032 & 421.816  & 10.68  & 11.9  & 185   & 6.7     & I      & 270$\pm$59  \\
    J1645-0317 & 387.690  & 35.76  & 11.1  & 185   & 40.4    & I      & 2224$\pm$107  \\
    J1709-1640 & 653.054  & 24.89  & 3.9   & 185   & 21.8    & I      & 271$\pm$18  \\
    J1711-5350 & 899.233  & 106.10  & 1.9   & 185   & 6.1    & I      & 20$\pm$15  \\
    J1731-4744 & 829.829  & 123.06  & 5.6   & 185   & 35.3   & I      & 424$\pm$8  \\
    J1735-0724 & 419.335  & 73.51  & 11.9  & 185   & 5.6     & C      & 25$\pm$11  \\
    J1741-0840 & 2043.082  & 74.90  & 10.6  & 185   & 7.0    & C      & 19$\pm$9  \\
    J1743-1351 & 405.337  & 116.30  & 12.6  & 185   & 7.2    & C      & 34$\pm$12  \\
    J1748-1300 & 394.133  & 99.36  & 11.1  & 185   & 9.4     & C      & 34$\pm$11  \\
    J1751-4657 & 742.354  & 20.40  & 9.1   & 185   & 36.8    & I      & 411$\pm$10  \\
    J1752-2806 & 562.558  & 50.37  & 2     & 185   & 143.1   & I      & 2441$\pm$107  \\
    J1801-0357 & 921.493  & 120.37  & 5.5   & 185   & 6.4    & C      & 12$\pm$6  \\
    J1807-0847 & 163.727  & 112.38  & 4.8   & 185   & 8.1    & C      & 25$\pm$7  \\
    J1819-1318 & 1515.696  & 35.10  & 7.7   & 185   & 10.3   & C      & 13$\pm$6  \\
    J1820-0427 & 598.082  & 84.44  & 1.9   & 185   & 197.7   & C      & 946$\pm$166  \\
    J1823+0550 & 752.907  & 66.78  & 11.5  & 185   & 13.9    & C      & 42$\pm$16  \\
    J1823-3106 & 284.054  & 50.24  & 4.6   & 185   & 10.8    & I      & 105$\pm$15  \\
    J1825-0935 & 769.021  & 19.38  & 19.2  & 185   & 13.7    & I      & 612$\pm$40  \\
    J1833-0338 & 686.733  & 234.54  & 3.5   & 185   & 5.1    & C      & 86$\pm$12  \\
    J1834-0010 & 520.954  & 88.65  & 6.3   & 185   & 8.5     & C      & 37$\pm$9  \\
    J1834-0426 & 290.108  & 79.31  & 3.3   & 185   & 51.2    & C      & 341$\pm$61  \\
    J1839-0627 & 484.914  & 92.49  & 4.3   & 185   & 10.1    & C      & 30$\pm$8  \\
    J1847-0402 & 597.809  & 141.98  & 6.5   & 185   & 8.6    & C      & 67$\pm$11  \\
    J1849-0636 & 1451.319  & 148.17  & 6.7   & 185   & 8.7   & C      & 66$\pm$12  \\
    J1851-0053 & 1409.065  & 24.00  & 8.6   & 185   & 11.4   & C      & 18$\pm$7  \\
    J1900-2600 & 612.209  & 37.99  & 6.9   & 185   & 84.4    & I      & 393$\pm$11  \\
    J1901+0156 & 288.219  & 105.39  & 12.4  & 185   & 11.3   & C      & 40$\pm$10  \\
    J1902-5105 & 1.742  & 36.25  & 12.9  & 185   & 3.0       & I      & 362$\pm$55  \\
    J1909+1102 & 283.642  & 149.98  & 10    & 185   & 26.4   & C      & 206$\pm$31  \\
    %J1910-0112 & 1360.603  & 178.00  & 12.8  & 185   & 6.9   &       & 24.4$\pm$8.9  \\
    J1910-0309 & 504.606  & 205.53  & 12.3  & 185   & 11.3   & C      & 65$\pm$12  \\
    J1913-0440 & 825.936  & 89.39  & 12    & 185   & 103.9   & C      & 445$\pm$165  \\
    J1916+0951 & 270.254  & 60.95  & 10.4  & 185   & 17.7    & C      & 38$\pm$11  \\
    J1917+1353 & 194.631  & 94.54  & 6.5   & 185   & 53.4    & C      & 174$\pm$37  \\
    J1921+1948 & 821.035  & 153.85  & 2.1   & 185   & 12.4   & C      & 70$\pm$12  \\
    J1921+2153 & 1337.302  & 12.44  & 4.3   & 185   & 108.0  & I      & 2112$\pm$49  \\
    J1922+2110 & 1077.924  & 217.09  & 2.2   & 185   & 25.5  & C      & 67$\pm$15  \\
    J1932+1059 & 226.519  & 3.18  & 9.3   & 185   & 26.7     & I      & 362$\pm$15  \\
    J1935+1616 & 358.738  & 158.52  & 6.6   & 185   & 10.2   & I      & 106$\pm$9  \\
    J1939+2134 & 1.558  & 71.02  & 2.9   & 185   & 10.2      & C      & 93$\pm$14  \\
    J1943-1237 & 972.429  & 28.92  & 5.7   & 185   & 12.5    & I      & 68$\pm$8  \\
    J1946+1805 & 440.618  & 16.14  & 4.3   & 185   & 17.4    & C      & 81$\pm$15  \\
    J2018+2839 & 557.953  & 14.20  & 1.2   & 185   & 16.7    & I      & 561$\pm$41  \\
    %J2022+2854 & 343.402  & 24.63  & 2     & 185   & 10.6  & I      & 18.2$\pm$7.2  \\
    J2046-0421 & 1546.938  & 35.80  & 14.6  & 185   & 11.0   & I      & 34$\pm$7  \\
    J2053-7200 & 341.336  & 17.30  & 92.8  & 185   & 34.5    & C      & 122$\pm$23  \\
    J2219+4754 & 538.469  & 43.50  & 103.1 & 185   & 17.3    & I      & 871$\pm$58  \\
    J2253+1516 & 792.236  & 29.20  & 3.3   & 155   & 9.8     & C      & 10$\pm$5  \\
    J2256-1024 & 2.295  & 13.78  & 2.7   & 155   & 4.2       & C      & 67$\pm$20  \\
    J2346-0609 & 1181.463  & 22.50  & 4.1   & 185   & 9.0    & C      & 8$\pm$4  \\
    \hline
\end{longtable}
\end{ThreePartTable}
\end{center}